\def\year{2021}\relax
\documentclass[letterpaper]{article} % DO NOT CHANGE THIS
\usepackage{aaai21}  % DO NOT CHANGE THIS
\usepackage{times}  % DO NOT CHANGE THIS
\usepackage{helvet} % DO NOT CHANGE THIS
\usepackage{courier}  % DO NOT CHANGE THIS
\usepackage[hyphens]{url}  % DO NOT CHANGE THIS
\usepackage{graphicx} % DO NOT CHANGE THIS
\usepackage{graphicx}  % DO NOT CHANGE THIS
\usepackage{natbib}  % DO NOT CHANGE THIS AND DO NOT ADD ANY OPTIONS TO IT
\usepackage{caption} % DO NOT CHANGE THIS AND DO NOT ADD ANY OPTIONS TO IT
\usepackage{cite} 
\usepackage{float}
\setcitestyle{authoryear,round}

\title{Exploring the parameter reusability of CNN}
\author{
Wei Wang, Lin Cheng, Yanjie Zhu, Dong Liang
}

\begin{document}

\maketitle

\begin{abstract}
In recent times, using small data to train networks has become a hot topic in the field of deep learning.  Reusing pre-trained parameters is one of the most important strategies to address the issue of semi-supervised and transfer learning. However, the fundamental reason for the success of these methods is still unclear. In this paper, we propose a solution that can not only judge whether a given network is reusable or not based on the performance of reusing convolution kernels but also judge which layers’ parameters of the given network can be reused, based on the performance of reusing corresponding parameters and, ultimately, judge whether those parameters are reusable or not in a target task based on the root mean square error (RMSE) of the corresponding convolution kernels. Specifically, we define that the success of a CNN’s parameter reuse depends upon two conditions: first, the network is a reusable network; and second, the RMSE between the convolution kernels from the source domain and target domain is small enough. The experimental results demonstrate that the performance of reused parameters applied to target tasks, when these conditions are met, is significantly improved.
\end{abstract}

\section{Introduction}
As is known to all, the superior performance achieved by deep learning relies on the training on large-scale datasets, however it is a difficult challenge to collect enough labeled data with the limited resources available to most researchers. Thus, we are motivated to establish effective approaches to reduce the dependence upon labeled data. 

There are two popular ways to solve this problem:  

1. Deep	transfer learning: The reuse of a part of the pre-trained network of the original domain, including its network structure and parameters, as a part of the deep neural network used in the target domain. 

2. Semi-supervised learning: It uses both data with and without labels for learning. 

Essentially, both deep transfer learning and semi-supervised deep learning are examples involving parameter reuse. Semi-supervised learning can be considered as a special kind of transfer learning and can be called self-transfer learning. Parameter reuse is an approach that can give a good starting point for the target task. The usual transfer learning approach is to train a base network and then copy its first n layers to the first n layers of a target network. The usual semi-supervised learning approach is to share their first n layers on both unlabeled and labeled training data. These two approaches are based on general features, which are suitable to both base and target tasks; rather than features specific to the base task. However, general features are an abstract concept. Even so, general features do not equate to the features that can be reused. We argue that the first layer is not reusable in many networks. In short, there are three questions to be resolved for both approaches: first, whether the selected model is suitable for parameter reuse; second, which layers of the model are suitable for parameter reuse; and third, whether the given parameters can be reused in the current task.

To answer these questions, we propose to calculate the parameter reuse-root mean square error (PR-RMSE) as a strategy in order to first identify a reusable network (RN), subsequently to find the reusable layers of different kind of parameters of the network, and finally to judge the reusability of the given parameters. The strategy evaluates the reusability of networks by parameter reuse and evaluates the reusability of given parameters by computing the RMSE of two models’ convolution kernels. We define:

A1: The model trained for task 1 on dataset A.

A2: The model trained for task 2 on dataset A.

B1: The model trained for task 1 on dataset B.

B2: The model trained for task 2 on dataset B.

Assuming that $A1$ is our target task. Dataset A is similar to dataset B. There are three ways to reuse parameters for $A1: A2->A1, B1-> A1, B2-> A1$. First, we need to determine whether the network is a reusable network. A model is randomly selected from A2, B1 and B2, and the convolution kernels of each layer of the model are respectively reused on A1. In other words, the convolution kernels in A1 are replaced layer by layer. We replace only one layer at a time, leaving the others unchanged. Subsequently, the results are evaluated directly on the validation set. If the results are almost unchanged after reusing some convolution kernels, then the network is a reusable network and those layers whose results are almost unchanged are the reusable convolution kernel layers. Next, we determine whether the convolution kernels of a given model are reusable. The convolution kernels’ RMSE between A2, B1, B2, and A1 are computed, respectively. If the RMSE is small for each layer, the convolution kernels of a given model can be reused. Vice versa. On the basis of the known convolution kernels of the model can be reused. Parameter reusability experiments are carried out on other parameters of the Conv layer and BN layer. The other parameters reuse experiments are carried out on each layer of each parameter, and the results are evaluated on the validation set. The layers with almost constant results identify the reusable layers of this parameter.

The parameter reuse from an auto-encoder network to a segmentation network is taken as an example. The experiment is carried out for the segmentation task, and the reusability of the parameters was evaluated by the change of the Dice values of the segmentation result. First, a segmentation model and an auto-encoder model are trained for the following comparative tests. Parameters of the corresponding layers of the segmentation task are replaced by parameters of the image auto-encoding task, layer by layer. We do not retrain the model, however, simply test the model directly on the validation set and compare against the original results. We judge whether the network is an RN and get the reusable convolution kernels. Subsequently, the RMSE values of the corresponding reusable convolution kernels are calculated separately. If the RMSE values are small, the parameter can be reused, and vice versa. After testing the reusability of other parameters in the network layers, those having results with a small difference compared with the original results are deemed reusable, however the reuse of other parameters cannot be called an RN. 

The main contributions of this study can be summarized as follows:

(1). We propose PR-RMSE, a deep learning parameter reuse policy which aims to find RNs and reusable convolution kernels of RNs, and decide if the given pre-trained parameter is reusable or not.

(2). We analyze the batch-normalization based on mathematics and experiment.

(3). We give a reasonable explanation for RN.

\section{Related work}
In \cite{2014arXiv1411.1792Y}, the author proposes to use the generalization of the neural network for transfer learning. Two factors affecting the transfer learning performance are given: the middle of fragilely co-adapted layers and the specialization of higher layer features. Transfer learning is a method based on the generality of features \cite{Caruana,10.5555/3045796.3045800,Bengio2}. As features presume generality, we can first train a set of parameters based on one dataset and then transfer those to another dataset. Semi-supervised learning is a method of using both labeled data and unlabeled data when training data is insufficient. In \cite{Curriculum,Multi-Task,Semi-supervised}, the semi-supervised methods are used for image segmentation. 

At present, both deep transfer learning and semi-supervised deep learning based on the above theories have the same problem: the blindness toward the selection of parameters for reuse. Although \cite{2014arXiv1411.1792Y} try to explore the general network layers, every time they add a network layer, they retrain the model. This paper argues that this method cannot explain the generality of network layers, because the network has the adjustment ability. The cause of the results might be that the more layers set aside for network fine-tuning, the stronger the network fine-tuning capability, rather than the more generic layers at the front of the network. The reason why parameter reuse layers can be reused in \cite{Multi-Task,Semi-supervised} is not given. 

\begin{figure}[h]
\centering
\includegraphics[scale=0.18]{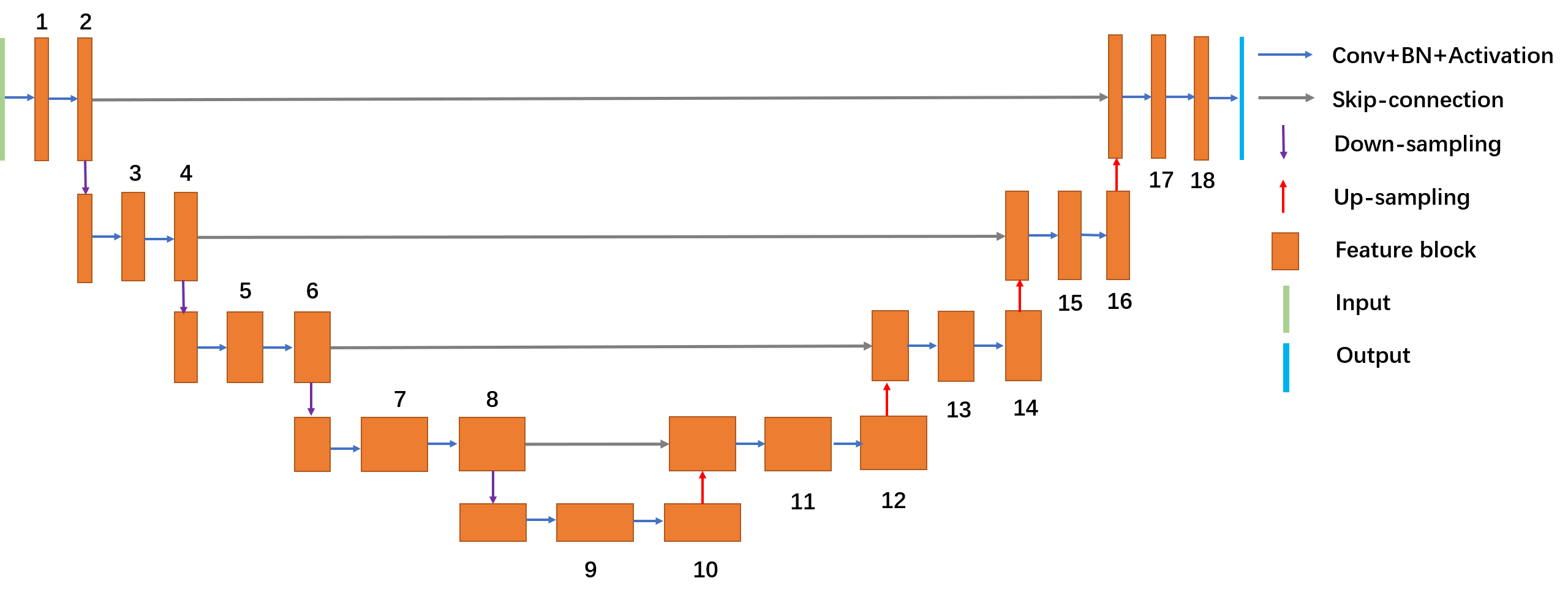}
\caption{Structure of UNet. The numbers denote the number of the layer. }
\label{unet}
\end{figure}

\begin{figure}[h]
\centering
\includegraphics[scale=0.03]{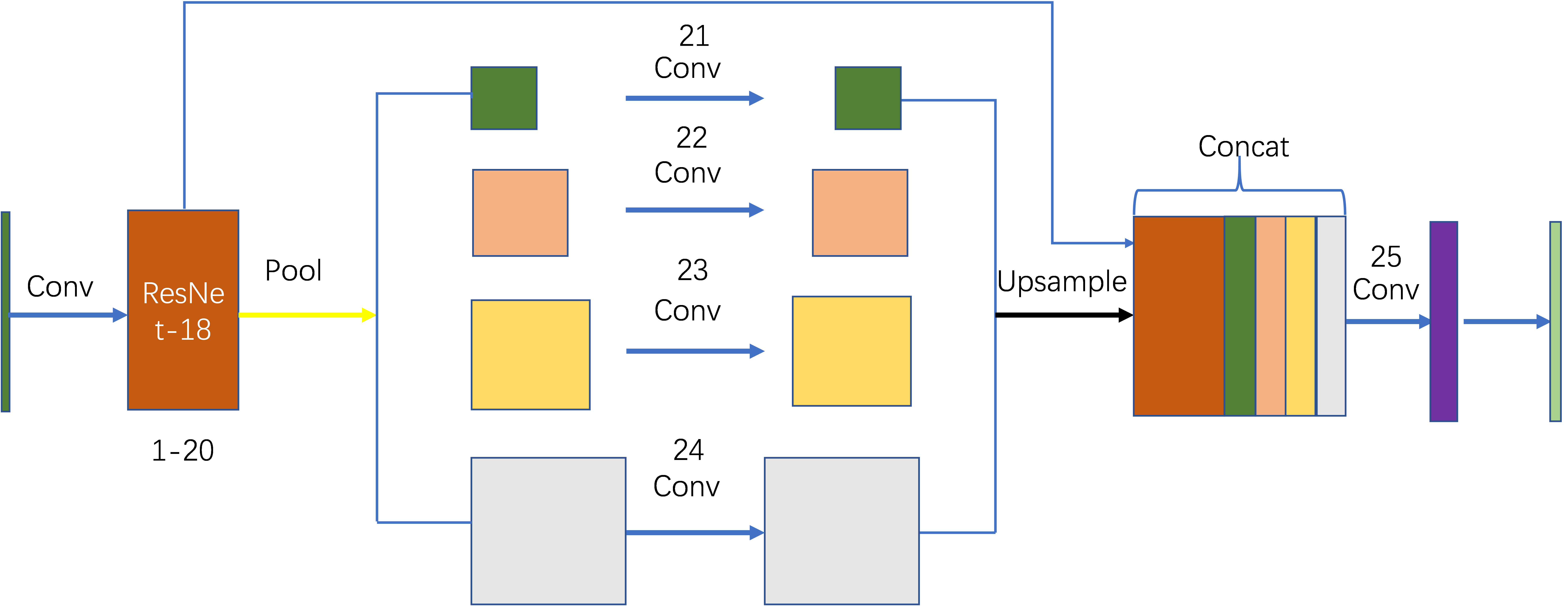}
\caption{Structure of PSPNet. The numbers denote the number of the layer. }
\label{pspnet}
\end{figure}

PSPNet \cite{psp} and UNet \cite{unet} were used in this paper. The network structure is as follows: Figures  \ref{unet} and  \ref{pspnet}.

\iffalse
\subsection{Article structure}
The structure of the article is as follows:
Mathematical representation of BN layers and differences of networks' convolution layers are introduced in Analysis of BN layer and convolution layer. 
	
The experiment is divided into three parts. The first part conducts a parameters reuse experiment and proves the validity of the formula and short-connection networks are RN in the section of Analysis of BN layer and convolution layer. The second part explores the differences between different datasets and different task network parameters. The third part carries on the parameter reuse experiment. 
\fi

\section{Analysis of BN layer and convolution layer}

\subsection{Parameter reuse and RMSE}
Parameter reuse can be mathematically expressed as:

$$
\left.\begin{array}{l}
G\left(\hat{x} ; \cdots ; \hat{\alpha}_{1}^{i} \cdots \hat{\alpha}_{n}^{i} ; \hat{\beta}_{1}^{i} \ldots \beta_{m}^{\hat{i}} ;\right. \\
\left.\cdots \hat{\alpha}_{1}^{i+1} \cdots \hat{\alpha}_{n}^{i+1} ; \hat{\beta}_{1}^{i+1} \ldots \hat{\beta}_{m}^{i+1} ; \cdots\right)=\hat{y} \\
F\left(x ; \cdots ; \alpha_{1}^{i} \cdots \alpha_{n}^{i} ; \beta_{1}^{i} \ldots \beta_{m}^{i} ; \ldots \alpha_{1}^{i+1} ;\right. \\
\left.\cdots \alpha_{n}^{i+1} ; \beta_{1}^{i+1} \cdots \beta_{m}^{i+1} ; \ldots\right)=y
\end{array}\right\} \stackrel{\hat{\alpha}_{1}^{i} \ldots \hat{\alpha}_{n}^{i}}{\longrightarrow}
$$

$$
F\left(x ; \cdots ;\hat{\alpha}_{1}^{i} \cdots \hat{a}_{n}^{i};\beta^i_1...\beta^i_m;...\alpha_{n}^{i+1};\beta^{i+1}_1...\beta^{i+1}_m;...\right)=y'
$$

$F:$ The function of target model.

$G:$ The function of source model.

$x:$ The input of  target model.

$\hat{x}:$ The input of source model.

$\alpha_{1}^{i} \cdots \alpha_{n}^{i};\beta^i_1...\beta^i_m;... \alpha_{1}^{i+1} \cdots \alpha_{n}^{i+1};\beta^{i+1}_1...\beta^{i+1}_m:$ Target model parameters.  

$\hat{\alpha}_{1}^{i} \cdots \hat{\alpha}_{n}^{i};\hat{\beta}^i_1...\hat{\beta}^i_m; \cdots \hat{\alpha}_{1}^{i+1} \cdots \hat{\alpha}_{n}^{i+1};\hat{\beta}^{i+1}_1...\hat{\beta}^{i+1}_m:$ Source model parameters.

$y:$The output of source model.

$y':$The output of parameter reuse model.

Make the following definition: The $W$ and $B$ are weight and bias terms of the convolution layer. RMSE can be mathematically expressed as:

$$
RMSE(W^i,\hat{W^i})
$$

$W^i:$ The $W$ of source model in layer i.

$\hat{W^i}:$ The $W$ of reused model in layer i.

\subsection{Analysis of the reuse of BN layer}

First, the BN of CNN can be defined as follow:

Input: Values of $x$ over a mini-batch: $B=\left\{x_{1 \ldots n}\right\}$ , $x_i = \{x_{i1 \ldots im}\}$, $m$ is the number of channels. Parameters to be learned: $w=\left\{w_{1 \ldots m}\right\}, b=\left\{b_{1 \ldots m}\right\}$. 

Output: $\left\{y_{ij}=\mathrm{B} \mathrm{N}_{w_j,  b_j}\left(x_{ij}\right)\right\}$. 
$$
\mu_{j} \leftarrow \frac{1}{n} \sum_{i=1}^{n} x_{ij}\  \  \  //mini-batch\   mean  \\
$$
$$
\sigma_{j}^{2} \leftarrow \frac{1}{m} \sum_{i=1}^{m}\left(x_{ij}-\mu_{j}\right)^{2} \  \  \   
$$
$$
\widehat{x}_{i} \leftarrow \frac{x_{ij}-\mu_{j}}{\sqrt{\sigma_{j}^{2}+\epsilon}}\  \  \  
$$
$$
y_{ij} \leftarrow w_j \widehat{x}_{ij}+b_j \equiv \mathrm{B} \mathrm{N}_{w_j, b_j}\left(x_{ij}\right)  \  \  \  
$$   

$\mu_{j}$:The mean of channel j. 

$\sigma_{j}^{2}$:The variance of channel j. 

We define:

RM:\{$\mu=\mu_{1}. . . \mu_{n}$\}  

RV:\{$\sigma^{2} = \sigma_{1}^{2}. . . \sigma_{n}^{2}$\}

RW:\{$w = {w_1. . . w_n}$\}

RB:\{$b={b_1. . . b_n}$\}

 $\epsilon:$ $\epsilon>0$ is a very small value. 
 
 Change the parameters RM, RV, RW, RB of the BN layer, the BN formula can be written as follows: 

{\bfseries RM}
$$
\begin{array}{l}
\hat{x}_{ij}=\frac{x_{ij}-\left(\mu_{j} \pm \Delta \mu_{j}\right)}{\sqrt{\sigma_{j}^{2}+\varepsilon}}\left(\Delta \mu_{j}=\mu_{j}-\mu_{j}^{\prime}\right) \\
\hat{x}_{ij}=\frac{x_{i}-\mu_{j}}{\sqrt{\sigma_{j}^{2}+\varepsilon}} \pm \frac{\Delta \mu_{j}}{\sqrt{\sigma_{j}^{2}+\varepsilon}} \\
\hat{y}_{ij}=w_j \frac{x_{ij}-\mu_{j}}{\sqrt{\sigma_{j}^{2}+\varepsilon}} \pm w_j \frac{\Delta \mu_{j}}{\sqrt{\sigma_{j}^{2}+\varepsilon}}+b_j
\end{array}
$$

{\bfseries RV}
$$
\begin{array}{l}
\hat{x}_{ij}=\frac{x_{ij}-\mu_{j}}{\sqrt{\sigma_{j}^{2} \pm \Delta\sigma_j^{2}+\varepsilon}} \quad\left(\Delta \sigma_{j}^{2}=\sigma_{j}^{2}-\sigma_{j}^{\prime 2}\right) \\
\hat{x}_{ij}=\frac{x_{ij}-\mu_{j}}{\sqrt{\sigma_{j}^{2}+\varepsilon}} \cdot \frac{\sqrt{\sigma_{j}^{2}+\varepsilon}}{\sqrt{\sigma_{j}^{2} \pm \Delta \sigma_{j}^{2}+\varepsilon}} \\
\hat{y}_{ij}=w_j \frac{x_{ij}-\mu_{j}}{\sqrt{\sigma_{j}^{2}+\varepsilon}} \frac{\sqrt{\sigma_{j}^{2}+\varepsilon}}{\sqrt{\sigma_{j}^{2} \pm \Delta \sigma_{j}^{2}+\varepsilon}}+b_j
\end{array}
$$

{\bfseries RW}
$$
\hat{y}_{ij}=\alpha_j w_j \hat{x}_{ij}+b_j \quad \quad\left(\alpha_j=\frac{w_j^{\prime}}{w_j}\right)
$$

{\bfseries RB}
$$
\hat{y}_{ij}=w_j \hat{x}_{ij}+b_j \pm \Delta b_j \quad\left(\Delta b_j=b_j-b_j^\prime\right)
$$

where $\mu_j^\prime, \sigma_{j}^{\prime 2}, w_j^{\prime}, b_j^\prime$ are the reused parameters in layer j. $\hat{x}_{ij}, \hat{y}_{ij}$ denote the result after parameters reuse. 
We define:

MRM(Mean of RM):$\frac{1}{m}\sum_{j=1}^{m}w_j\frac{\mathrm{\Delta}\mu_j}{\sqrt{\sigma_j^2+\varepsilon}}(\mathrm{\Delta}\mu_j = \sqrt{(\mu_j-\mu_j^\prime)^2})$

MRV(Mean of RV):$\frac{1}{m}\sum_{j=1}^{m}\frac{\sqrt{\sigma_{j}^{2}+\varepsilon}}{\sqrt{\sigma_{j}^{2} \pm \Delta \sigma_{j}^{2}+\varepsilon}}$ 

MRW(Mean of RW):$\frac{1}{m}\sum_{j=1}^{m}\alpha_j$

MRB(Mean of RB):$\frac{1}{m}\sum_{j=1}^{m}\Delta b_j(\Delta b_j=\sqrt{(b_j-b_j^\prime)^2} )$

\subsection{Analysis of the parameter reuse of convolutional layer}

We believe that the features trained on different tasks on a similar dataset, same tasks on a similar dataset, and different tasks on the same dataset, are correlated. In other words, they learn the same or similar features. Subsequently, the convolution kernels are the same or similar, and hence, they are probably reusable. Furthermore, we believe that the reusability is different for different networks. At present, there are mainly two forms of CNN network unit. (1) short-connection, such as ResNet and DenseNet. (2) long-connection, such as UNet.

Take ResNet as an example. The convolution process can be written as follows:
$$
x_{l+1}=\sigma(x_l+BN(\sigma(BN(x_l*w_l'))*w_l''))
$$
After parameter reuse, it can be written as:
$$
\hat{x_{l+1}}=\sigma(x_l+BN(\sigma(BN(x_l*\hat{w_l'}))*w_l''))
$$
$$
\hat{x_{l+1}}=\sigma(x_l+BN(\sigma(BN(x_l*w_l'))*\hat{w_l''}))
$$

$x_l, x_{l+1}$:The input and output of the $lth$ residual unit. 

$w_l, w_{l+1}$:The first and the second convolution kernels of the $lth$ residual unit from the target model. 

$\hat{w_l}, \hat{w_{l+1}}$:The first and the second convolution kernels of the $lth$ residual unit from the reused model. 

BN:Batch normalization. 

$\sigma$:Relu. 

Long-connection is generally divided into encoder and decoder parts. 
The convolution process can be written as follows:
$$
x_{l+2}=\sigma(BN(\sigma(BN(x_l*w_l))*w_{l+1}))
$$
$$
x_{l+2}^D=\sigma(BN(\sigma(BN(C(x_l^E,x_l^D)*w_l^D))*w_{l+1}^D))
$$
After parameter substitution, it can be written as:
$$
\hat{x_{l+2}}=\sigma(BN(\sigma(BN(x_l*\hat{w_l}))*w_{l+1}))
$$
$$
\hat{x_{l+2}}=\sigma(BN(\sigma(BN(x_l*w_l))*\hat{w_{l+1}}))
$$
$$
\hat{x_{l+2}}^D=\sigma(BN(\sigma(BN(C(x_l^E, x_l^D)*\hat{w_l}^D))*w_{l+1}^D))
$$
$$
\hat{x_{l+2}}^D=\sigma(BN(\sigma(BN(C(x_l^E, x_l^D)*w_l^D))*\hat{w_{l+1}}^D))
$$

$x_l, x_{l+2}$:The $lth$ and $l+2th$ feature map. 

$x_l^E$:The $lt$h feature map in encoder part. 

$x_l^D, x_{l+2}^D$:$x_l^D$ has connection with $x_l^E$ in the decoder part. $x_{l+2}^D$ is the result of $x_l^E$ convolved twice. 

$\hat{}\   $:The result of applying parameter reuse. 

$w_l, w_{l+1}$:The $lth$ and $l+2th$ convolution kernel. 

$w_l^D, w_{l+1}^D$:The $lth$ and $l+2th$ convolution kernel in the decoder part. 

C:Concat feature maps in channel dimension. 

BN:Batch normalization. 

$\sigma$:Relu. 

We believe that short-connection forms are reusable, unlike long-connection forms. Short-connection structures, such as ResNet are a continuous feature decomposition and stored procedure. Its residual layers are responsible for adjusting the role of features for task learning; while identity mappings are responsible for preserving the learned information. Decomposition and storage of features in the whole process is the most important aspect. Because the identity mappings and feature number change layers are dominant, the residual layers are reusable. By contrast, the long-connection structure, with complete information transmission from front to back, is a process of continuous feature decomposition, thus each position is targeted to the position task and does not have reusability. 

For a short-connection network, how can we judge whether the given parameters are reusable or not? In our opinion, the difference between the corresponding convolution kernels should not be too large. In other words, $RMSE(w_l, \hat{w_l})$ should not be too large, because networks are sensitive to the change of convolution kernel, and we will reuse multi-layer convolution kernels. If the multi-layer difference is too large, error accumulation will be the result, which is not conducive to parameter reuse.

\section{Data}
Three datasets were used in this experiment, namely ACDC \cite{ACDC}, T1 \cite{T1} and RVSC \cite{RVSC}. 

The ACDC dataset is used to segment the left ventricle, right ventricle, and myocardium. Hundred people are considered in the experiment with 10-20 images of each person in the dataset. The dataset was randomly divided into two parts, a training set and a validation set, each containing 50 people. The T1 dataset was used to segment the myocardium. There are 210 people in total, with 55 images per person. The first 60 people were taken as the training set, and the last 30 people were taken as the validation set. There are three groups of RVSC: TrainingSet, Test1Set, and Test2Set, each contains 16 samples. In this experiment, only TrainingSet and Test1Set were used, with 242 and 262 images, respectively.

\section{Analysis of the parameter reuse of different networks}
\subsection{1. Parameter reuse for each layer of the network}

UNet and PSPNet were used to train the image auto-encoder and image segmentation tasks on the ACDC dataset. Because the number of channels in the segmentation result is 4, to ensure the correspondence of parameters, the image auto-encoder task also outputs 4 channels, each of their labels is the original image. The parameters of the image segmentation model were replaced by an image auto-encoder model, layer by layer. We evaluate the new model on the validation dataset directly, recording the corresponding Dice value and layer.

\iffalse
\begin{table}[h]
    \caption{ACDC:Result of UNet}
    \centering
    \begin{tabular}{|l|l|l|l|l}
        class & class-0 & class-1 & class-2 & class-3 \\ 
        Dice & 1. 00 & 0. 81 & 0. 83 & 0. 92 \\
    \end{tabular}
\end{table}
\begin{table}[h]
\caption{ACDC:Result of PSPNet}
    \centering
    \begin{tabular}{|l|l|l|l|l|}
        PSPNet-b & class-0 & class-1 & class-2 & class-3 \\ 
        yes & 1 & 0. 86 & 0. 86 & 0. 93 \\ 
        no & 1 & 0. 86 & 0. 86 & 0. 92 \\ 
    \end{tabular}
\end{table}
\fi

Because the pre-trained parameters of ResNet-18 were loaded before the training, there is no bias parameters in the convolutional layers of ResNet-18. Hence, only a part of the convolutional layers contains bias parameters. 
%In Table 2, yes, no denotes whether the convolutional layers of the models contain bias. 

\subsubsection{1.1 UNet}

\paragraph{1.1.1 Result of UNet}~{}

\begin{figure}[h]
\centering
\includegraphics[scale=0.05]{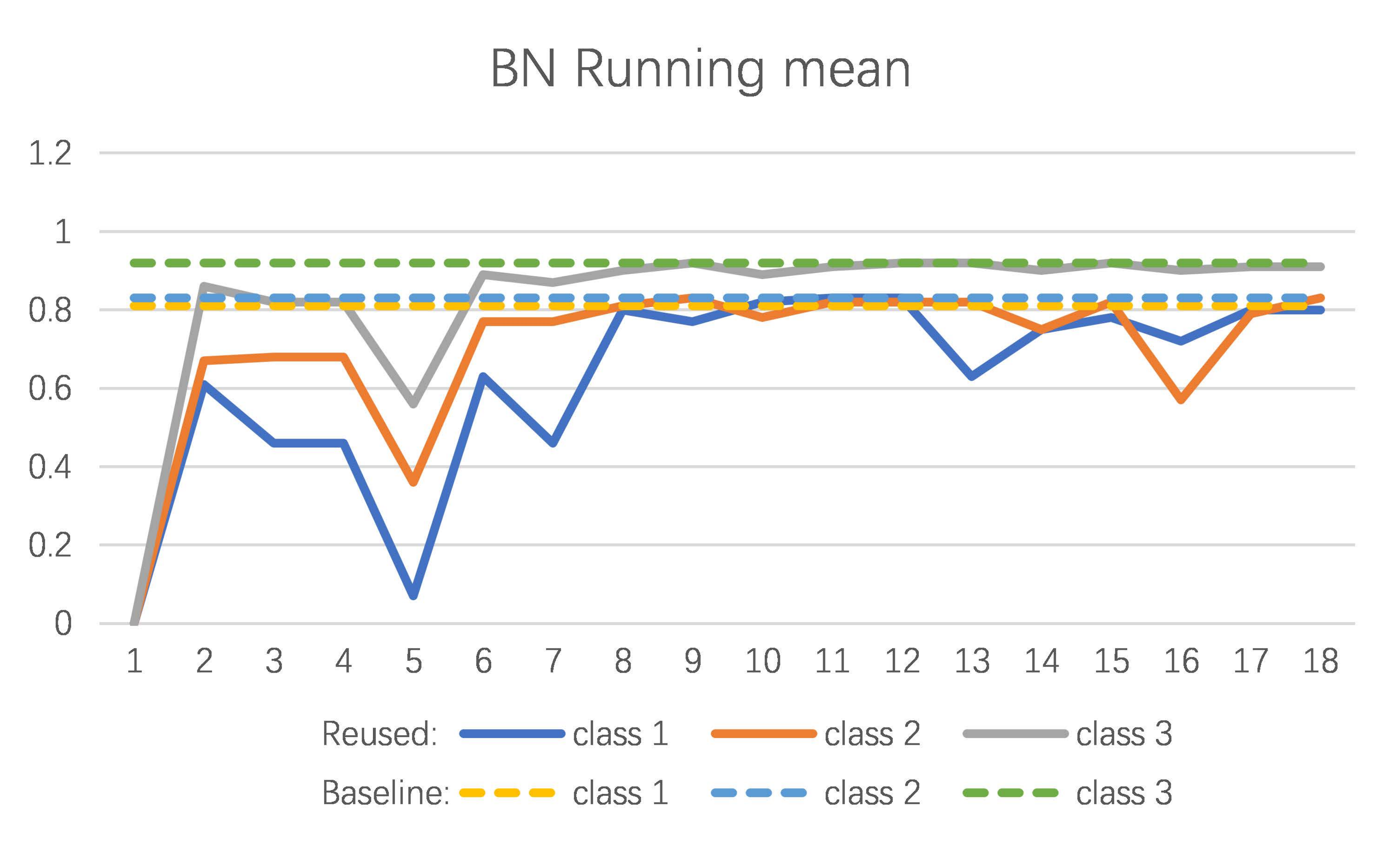}
\includegraphics[scale=0.05]{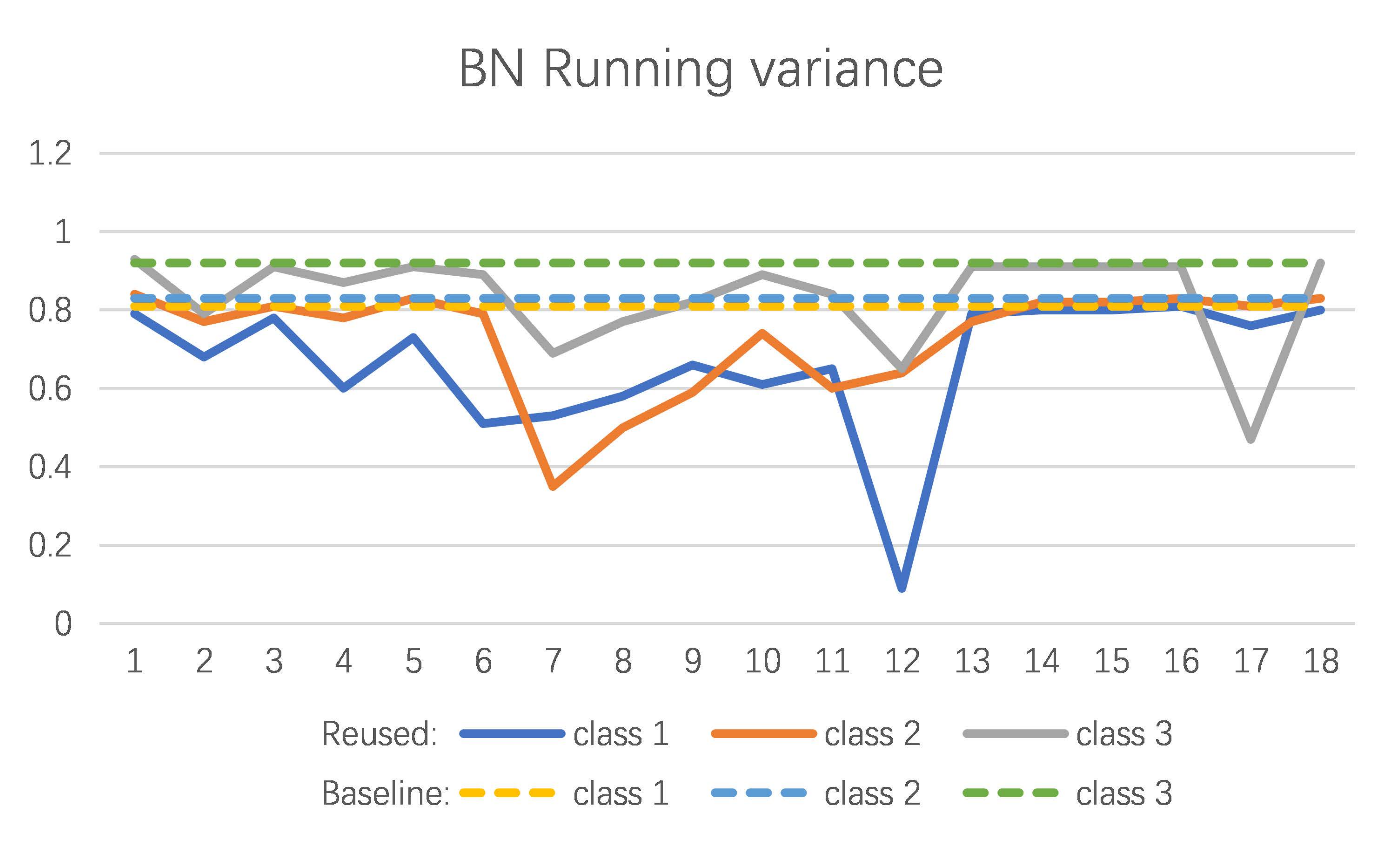}

\includegraphics[scale=0.05]{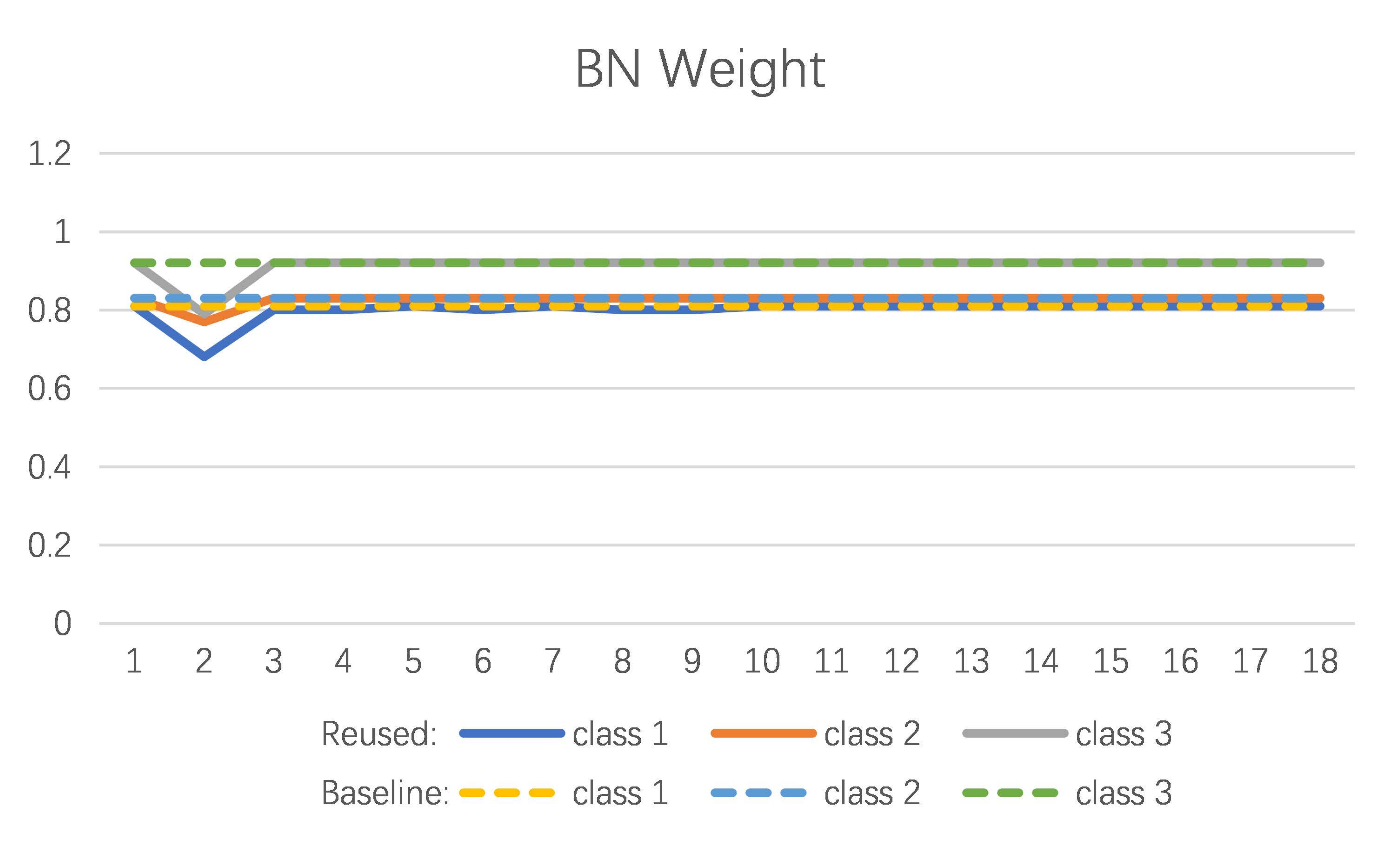}
\includegraphics[scale=0.05]{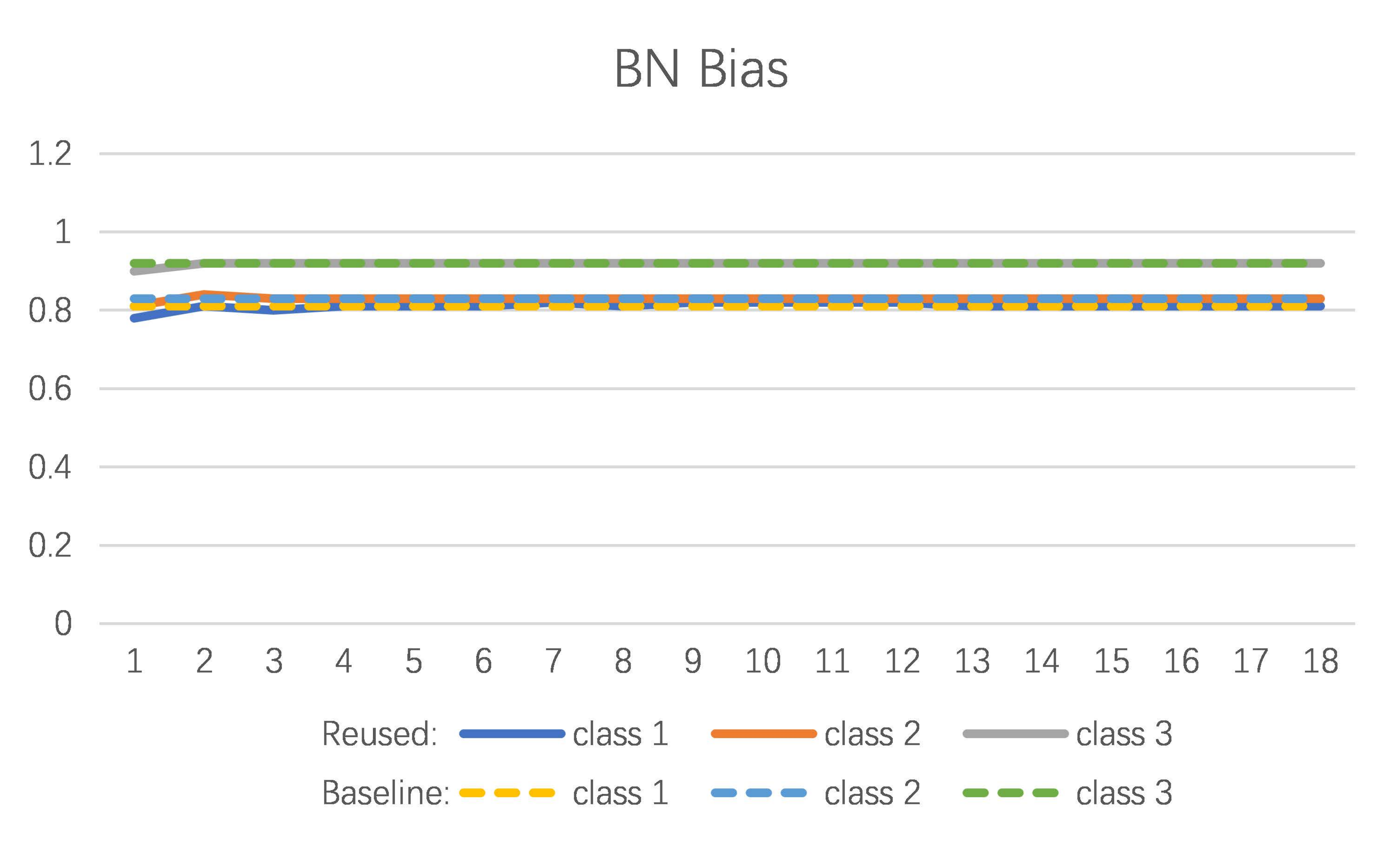}

\caption{UNet : The four images show the Dice value when the titled parameters were replaced,
and the abscissa denotes the layer to be replaced ( 1-18 corresponds to the BN layers of the
network from front to back). The ordinate denotes the corresponding Dice value.
}
\label{fig:unet-bn}
\end{figure}

The parameters of the BN layers and convolution layers of the image segmentation were replaced by the parameters of the corresponding layer of an image auto-encoder task, layer by layer, and the corresponding Dice values were obtained on the validation set, and the results are as follows: 

{\bfseries RM,RV,RW, and RB reuse results in BN layers:} Figure \ref{fig:unet-bn}.

The deeper the layer, the greater the reusability of the RM. RV reusability did not show obvious regularity. RW and RB are reusable.

\begin{figure}[h]
\centering
\includegraphics[scale=0.05]{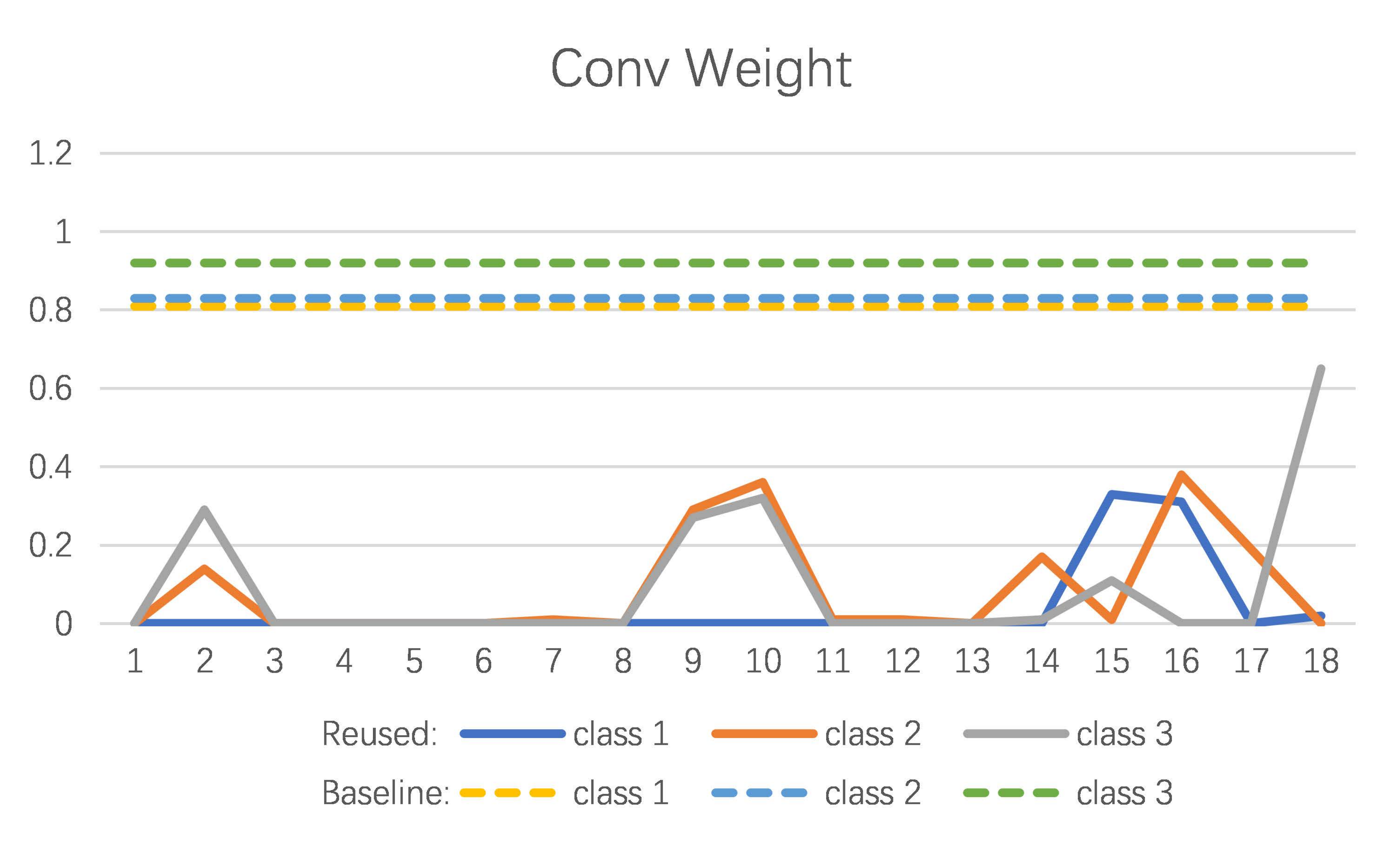}
\includegraphics[scale=0.05]{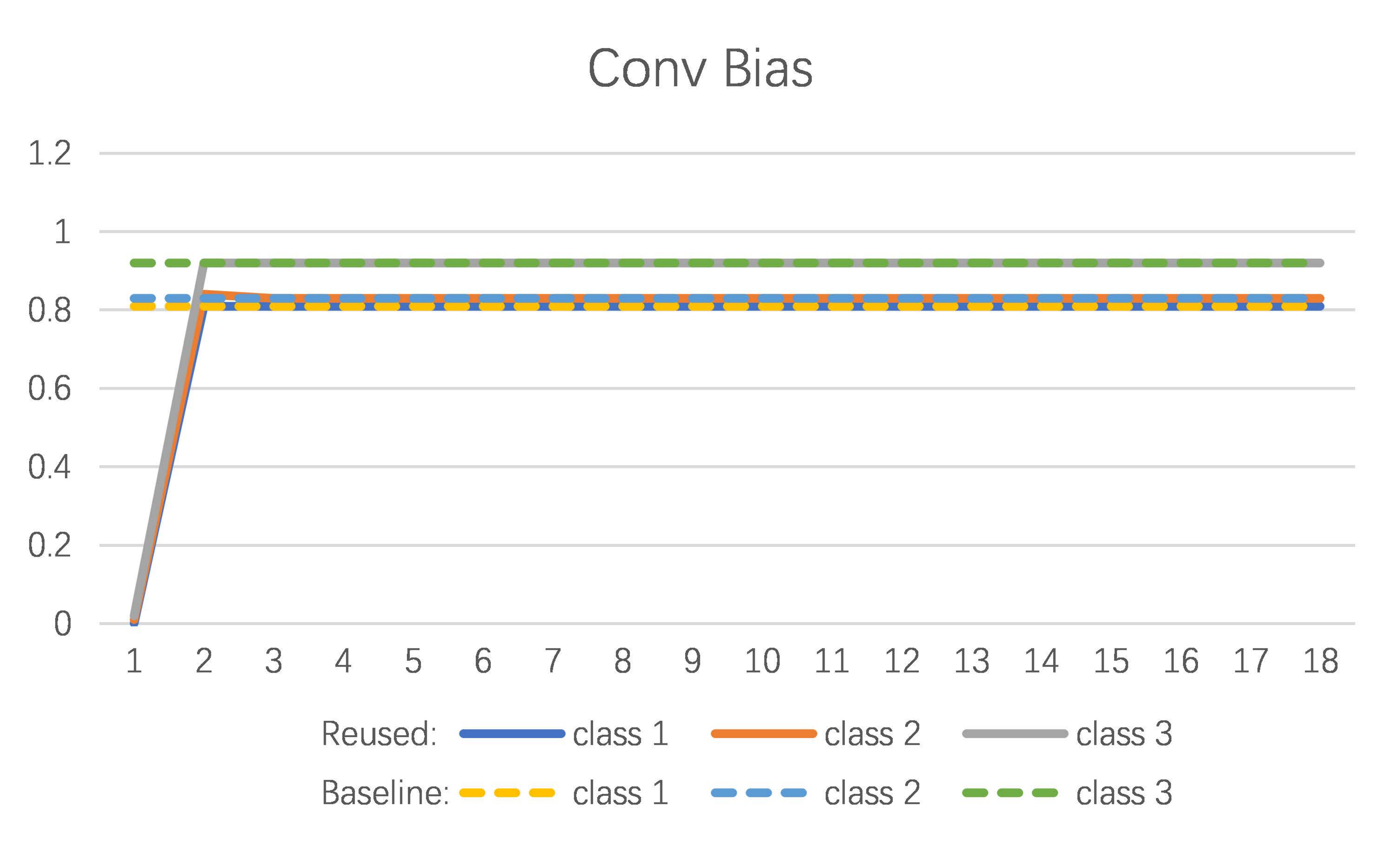}

\caption{UNet: Weight and bias terms in the figure correspond to kernel and bias of the convolution layer,
respectively. The title in the figure denotes the parameters of the convolution layer to be replaced. 
The abscissa denotes the layer to be replaced (1-18 corresponds to the convolution layer of the network
from front to back).
}
\label{unet-conv}
\end{figure}

{\bfseries W and B reuse results in convolution layers:} Figure \ref{unet-conv}.

W is not reusable, while B has reusability in all layers except the first layer.

\paragraph{1.1.2 Explore the reason of UNet result}~{}

{\bfseries Scale and shift of BN layers:} 

According to the results presented in Figure 3, the network segmentation results are more influenced by RM, RV than RW and RB. So let's assume that $MRM>>MRB, MRV>>MRW$. 

\begin{figure}[h]
\centering
\includegraphics[scale=0.25]{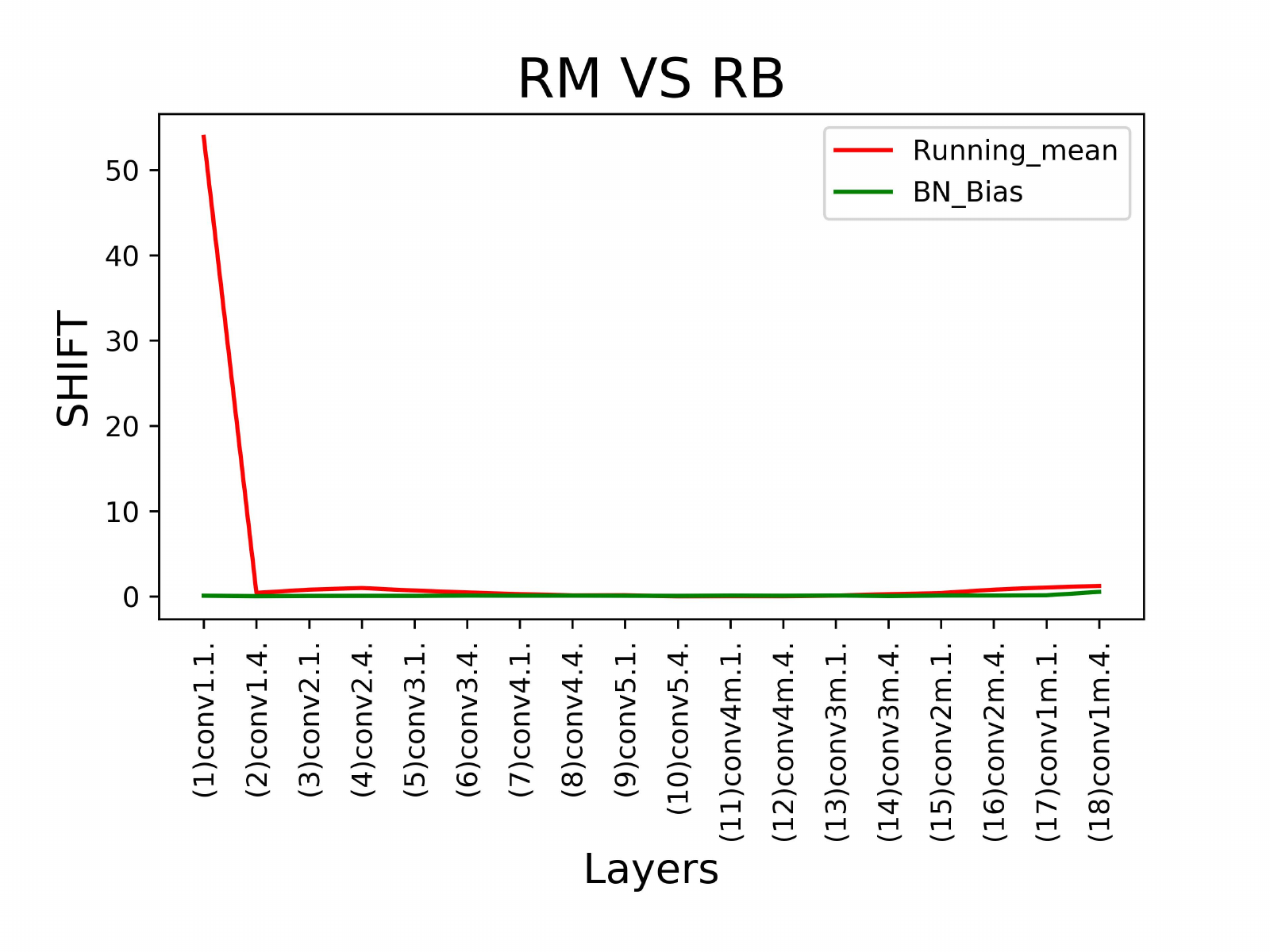}
\includegraphics[scale=0.25]{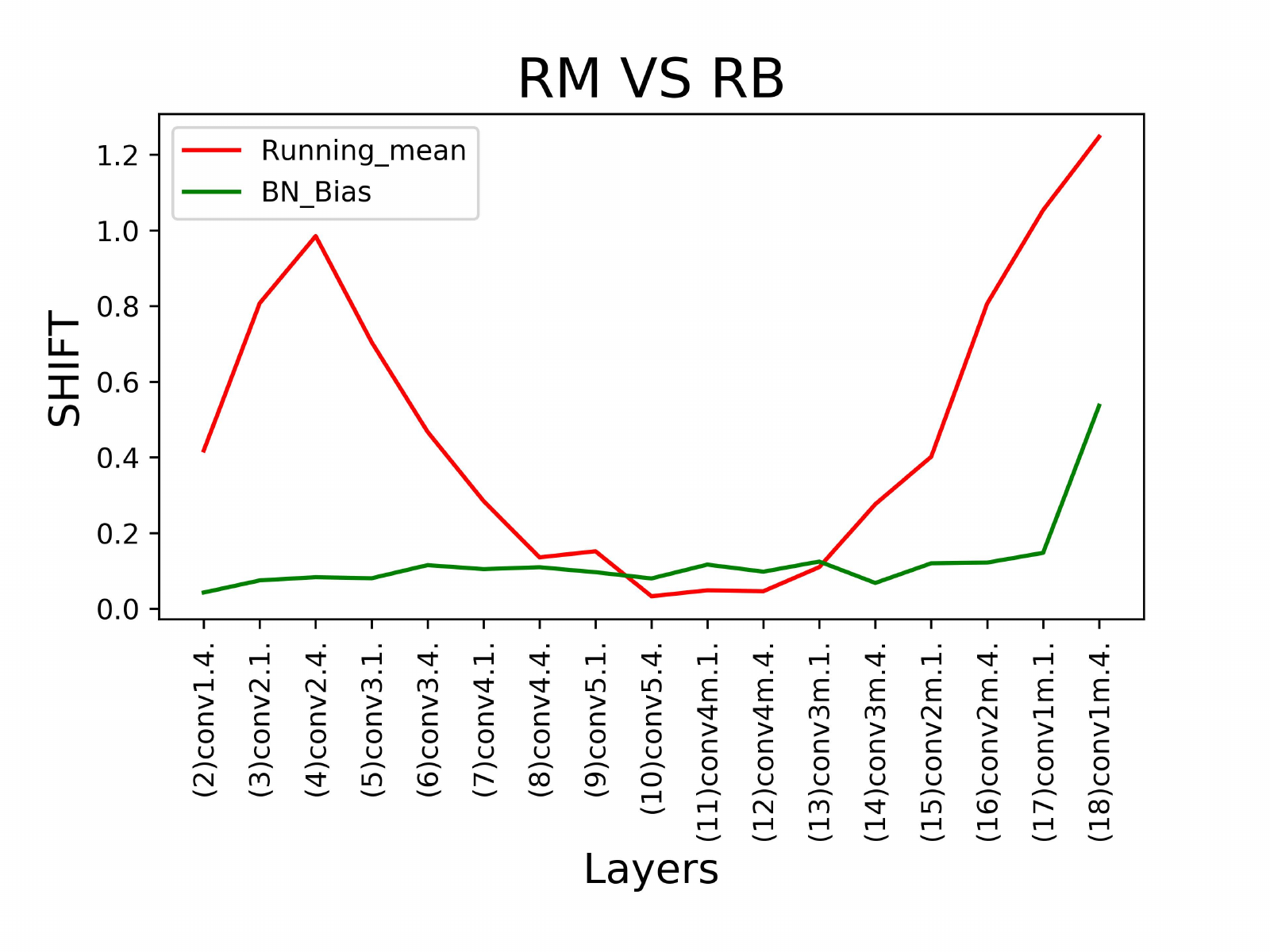}

\includegraphics[scale=0.25]{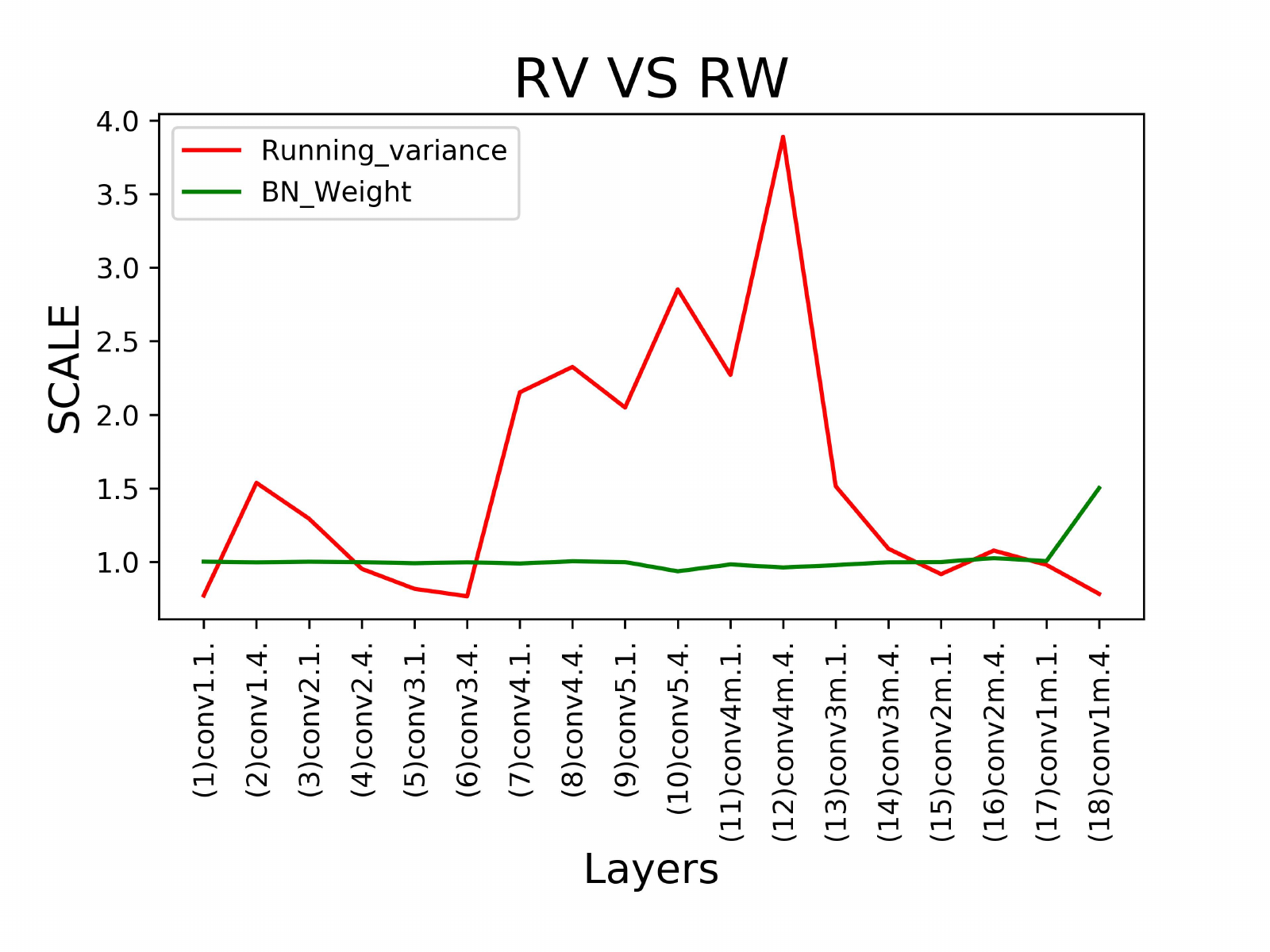}

\caption{UNet:Scale and shift in BN layers. Running Mean:MRM;BN Bias:MRB;Running variance:MRV;BN Weight:MRW (The top right image skipped the first layer of RM VS RV). 
}
\label{unet-scale-shitf}
\end{figure}

Plot the figure of $MRM, MRV, MRW, MRB$. Figure \ref{unet-scale-shitf}

Changes of MRM and MRB basically conform to the above reasoning in Figure 5. In addition, the deeper the layer, the lower the sensitivity of model results to MRM and MRB. Changes of MRV and MRW basically conform to the above reasoning for Figure 5.

{\bfseries RMSE of W and B in convolution layers:}  

\begin{figure}[h]
\centering
\includegraphics[scale=0.25]{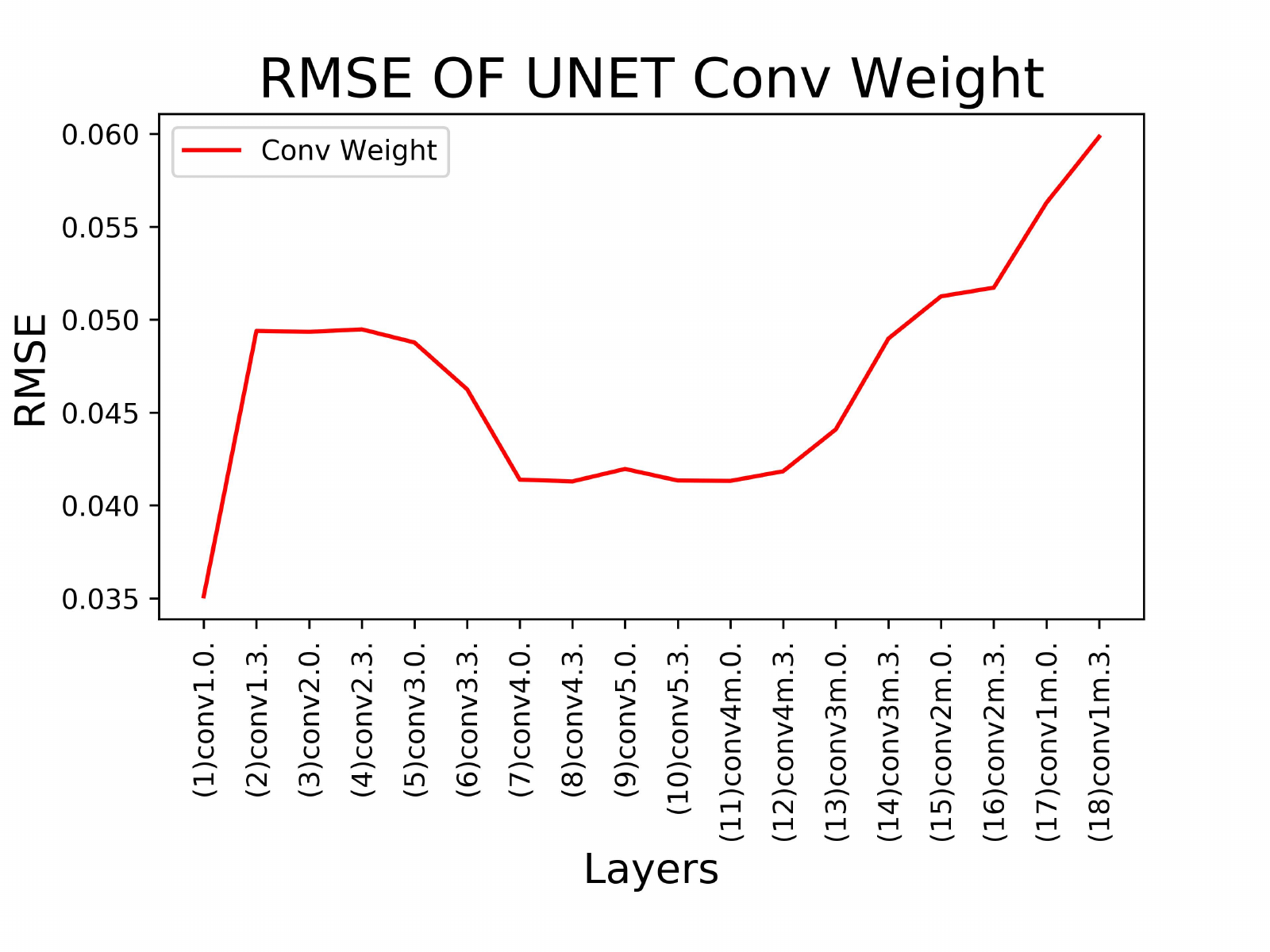}
\includegraphics[scale=0.25]{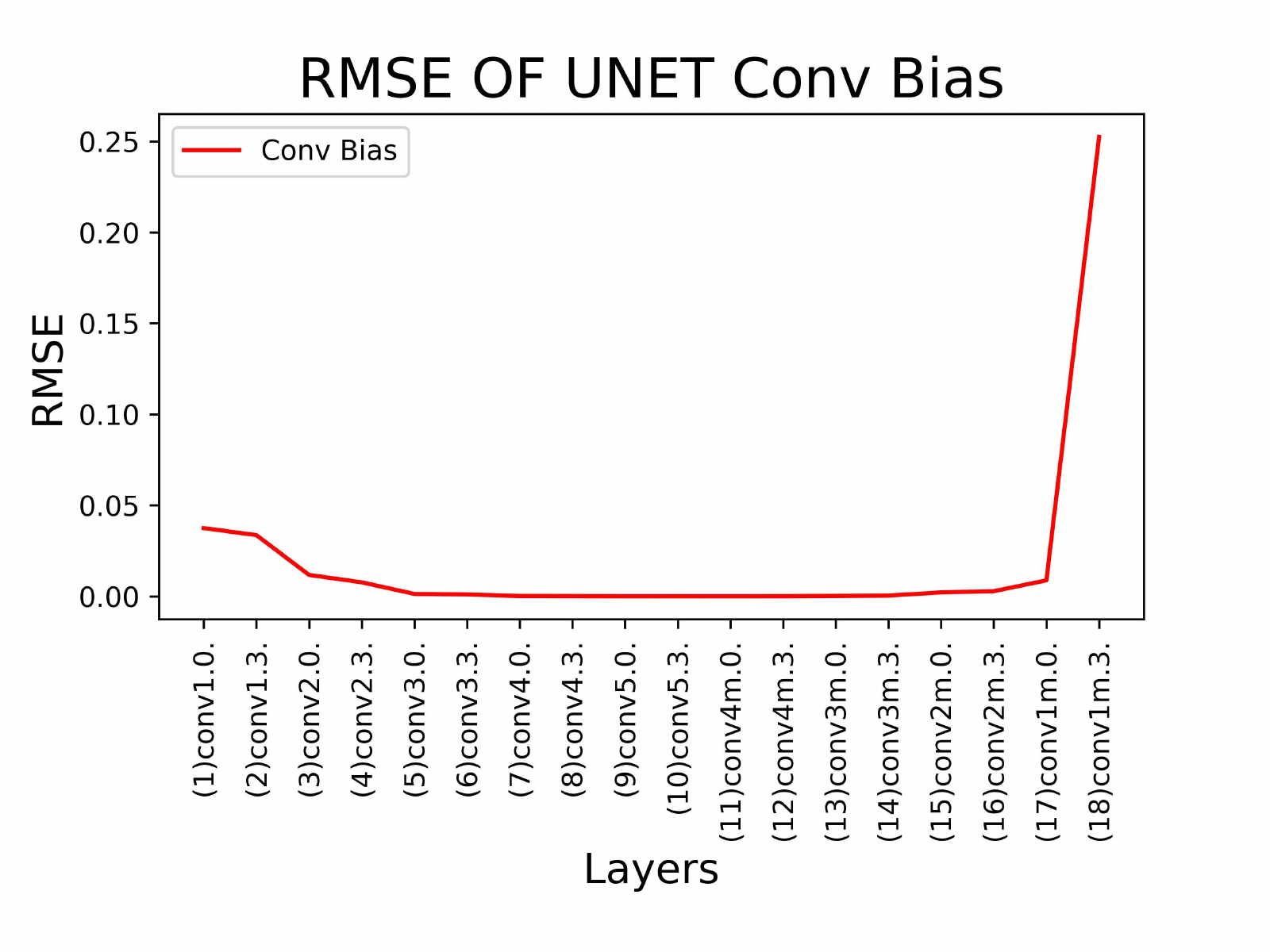}

\caption{UNet:The RMSE of W, B, which are the convolutional layers’ parameters.
}
\label{unet-conv-rmse}
\end{figure}

Figure 4 shows that the final result of the network in the convolution layers are greatly affected by W and slightly affected by B. In order to evaluate the differences in W and B between the two models in convolution layers, RMSE images were drawn for each layer. The results are as follows: Figure \ref{unet-conv-rmse}.

According to Figures 4 and 6, it is revealed that the network results are very sensitive to the change of W, and even a slight change will have a great impact on the results. However, the results are not sensitive to B. 

\subsubsection{1.2 PSPNet}~{}

Repeat the operation as section 1.1 on PSPNet.

\paragraph{1.2.1 Result of PSPNet}~{}

\begin{figure}[h]
\centering
\includegraphics[scale=0.05]{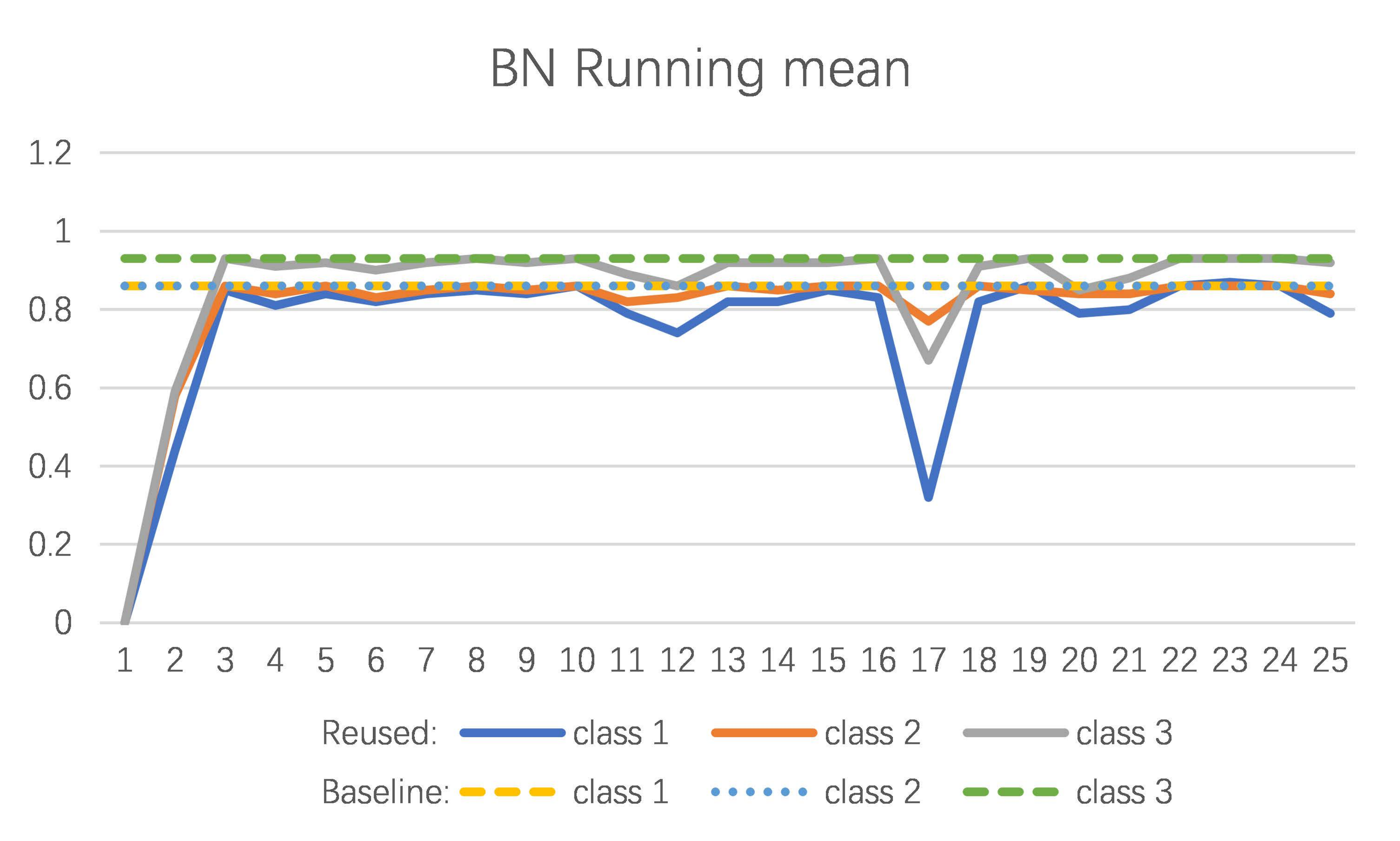}
\includegraphics[scale=0.05]{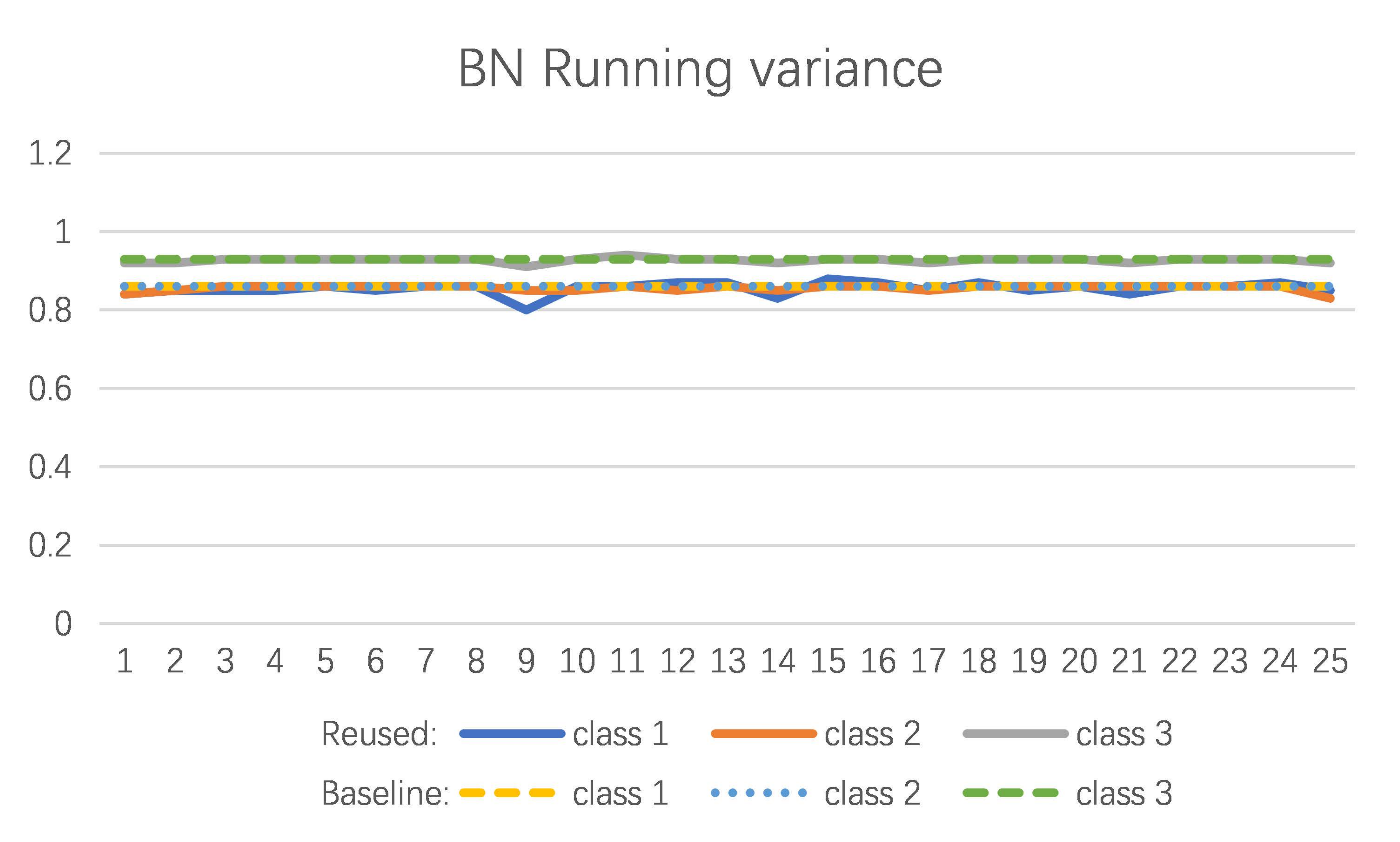}

\includegraphics[scale=0.05]{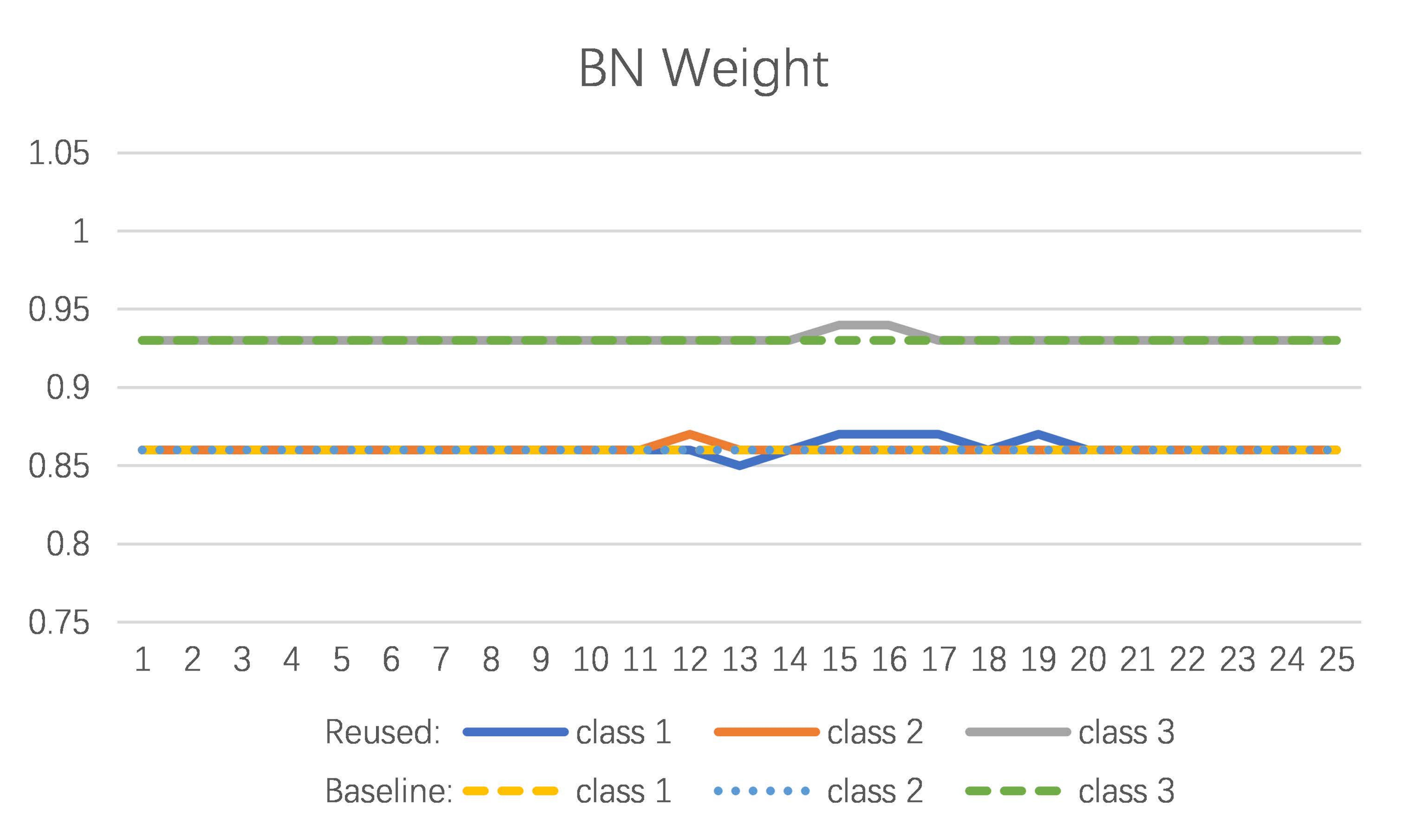}
\includegraphics[scale=0.05]{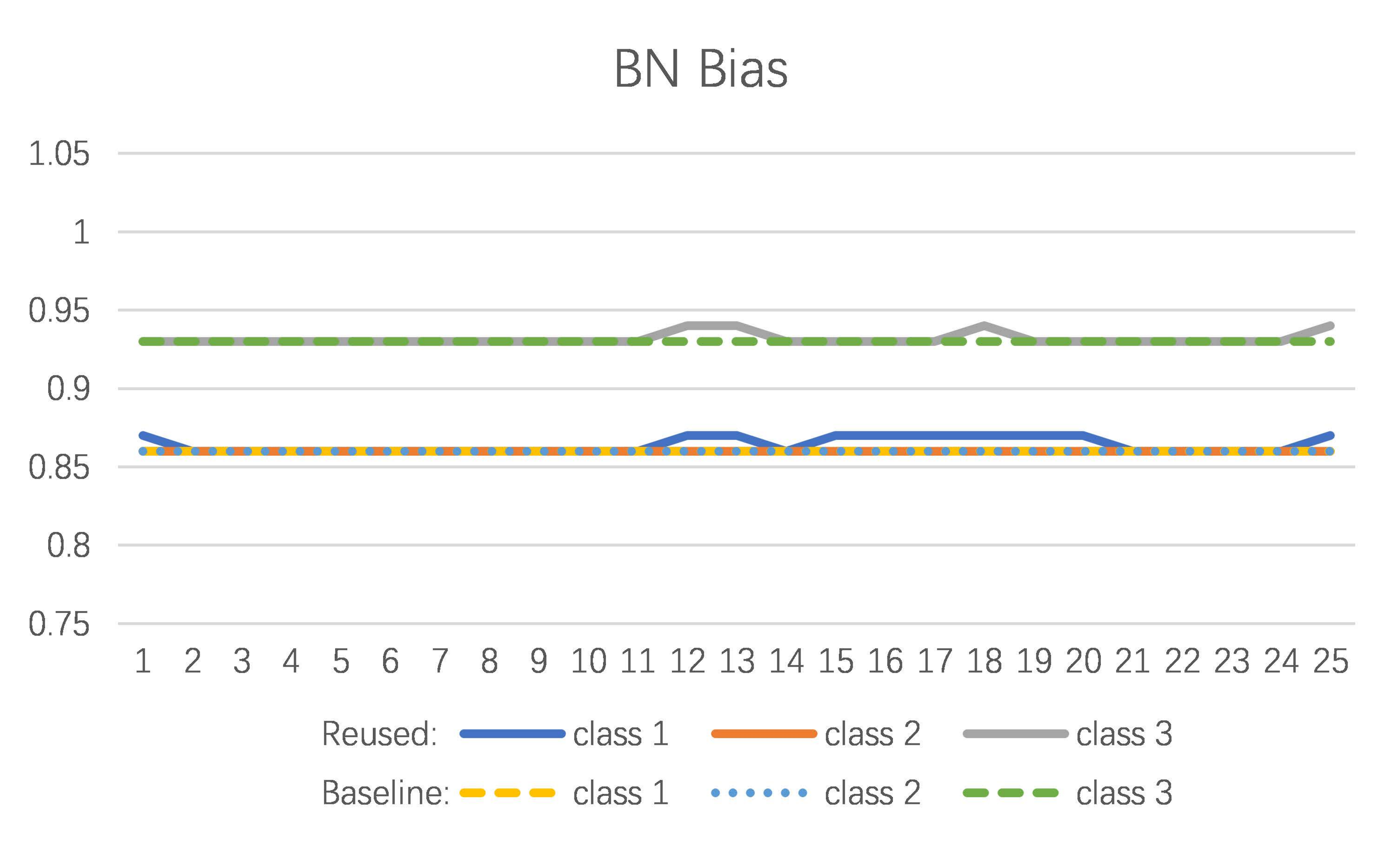}

\caption{PSPNet:The four images show the Dice value when the titled parameters were replaced,
and the abscissa denotes the layer to be replaced (1-25 corresponds to the BN layers of the network from front to back). The ordinate denotes the corresponding Dice value.}
\label{pspnet-bn}
\end{figure}

{\bfseries RM,RV,RW and RB reuse results in BN layers:} Figure \ref{pspnet-bn}.

RM is reusable in all layers except layers 1 and 17. RV, RW, and RB are reusable.

\begin{figure}[h]
\centering
\includegraphics[scale=0.05]{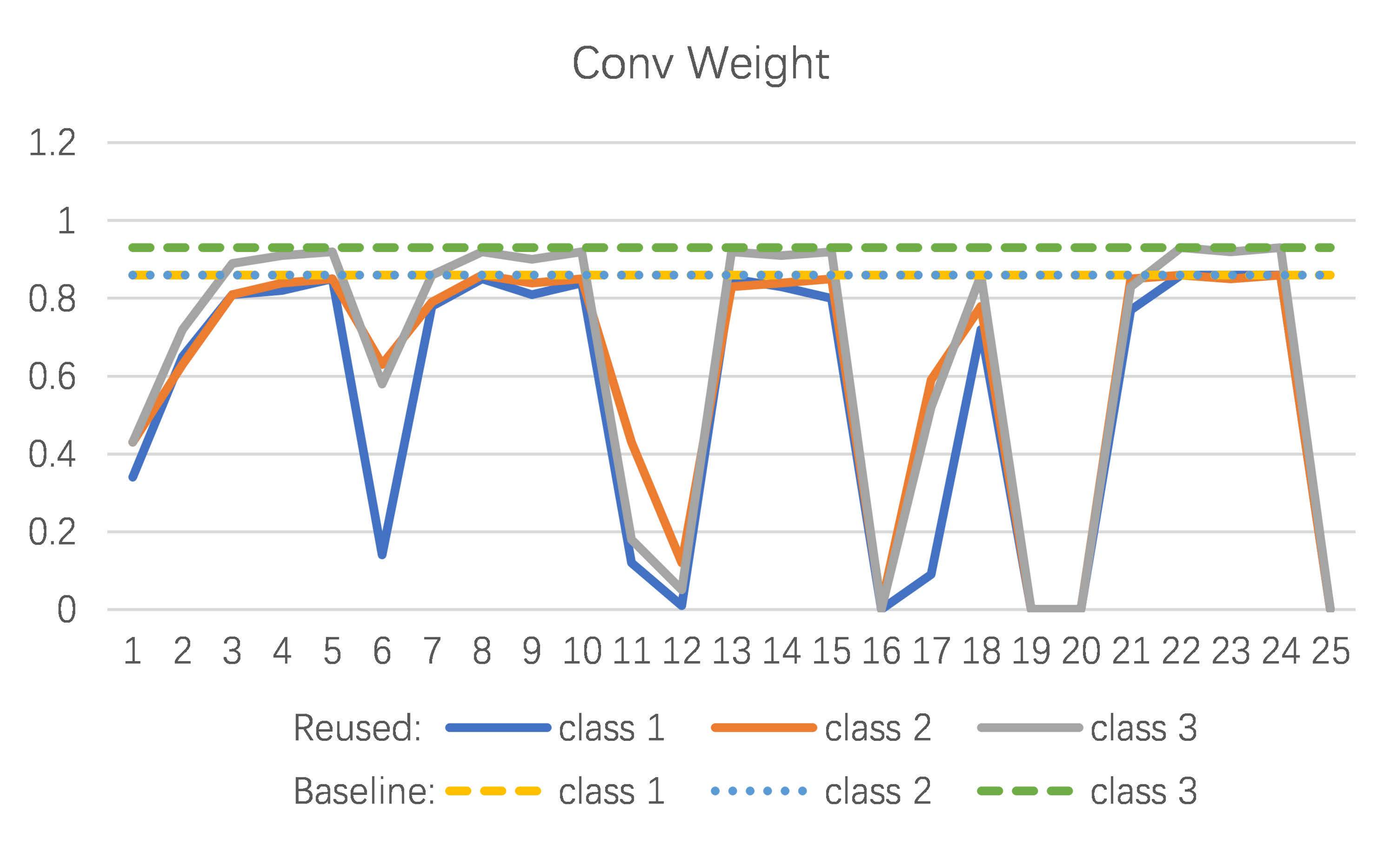}
\includegraphics[scale=0.05]{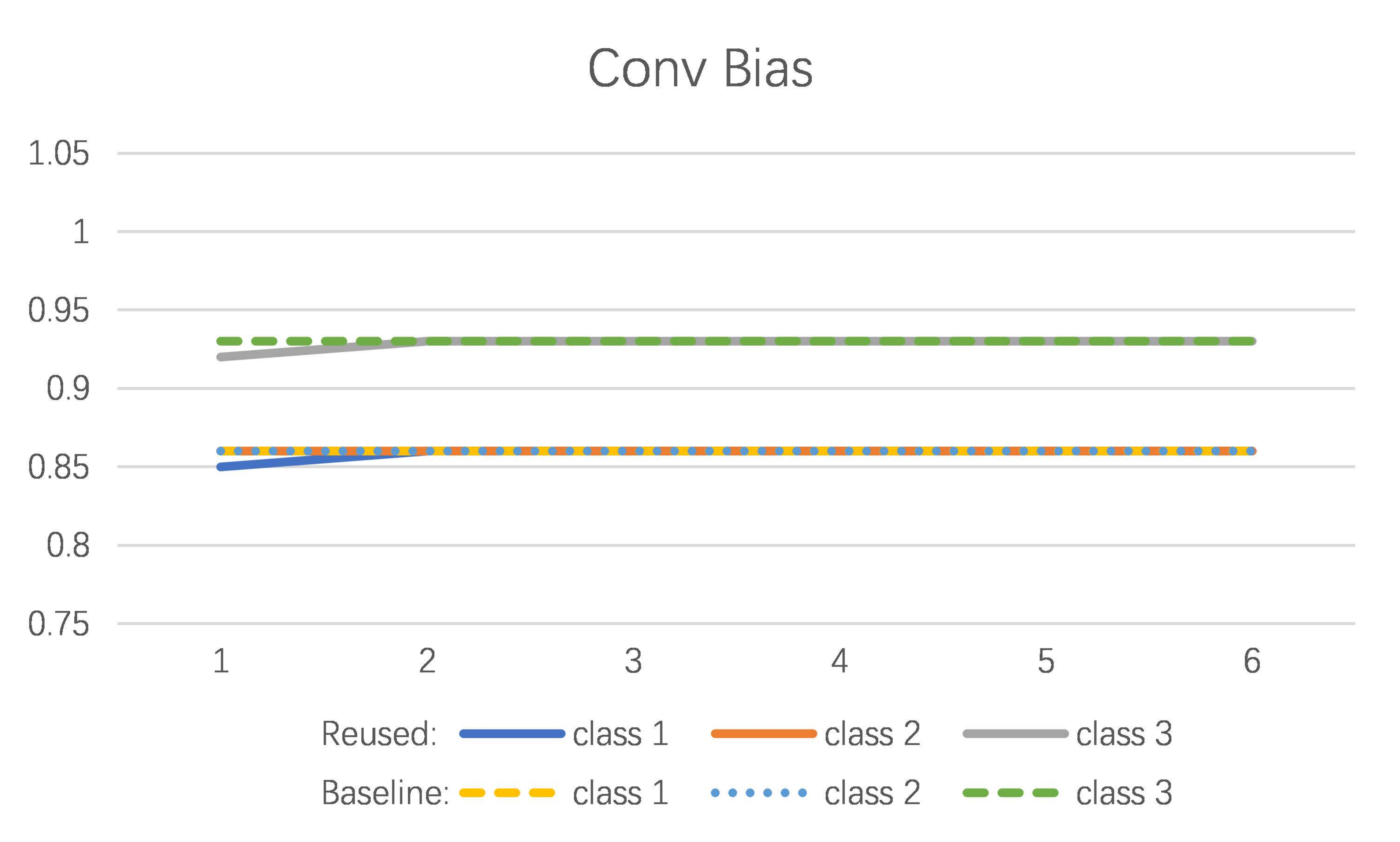}

\caption{PSPNet: Weight and bias terms in the figure correspond to kernel and bias in the convolution layer,
respectively. The title in the figure denotes the parameters of the convolution layer to be replaced, and
the abscissa denotes the layers to be replaced (1-25 corresponds to the convolution layer of the network
from front to back).
}
\label{pspnet-conv}
\end{figure}

{\bfseries W and B reuse results in convolution layers:} Figure \ref{pspnet-conv}.

Compared with Figure 4, PSPNet convolution kernels reuse shows strong regularity. Except the first layer, almost all non-reusable layers are the feature decomposition layer and the layer behind it. Hence, B is reusable.

\paragraph{1.2.2 Explore the reason of PSPNet results}~{}

{\bfseries Scale and shift of BN Layer:}   

According to the results shown in Figure 7, the network segmentation results are influenced more by RM and RV than RW and RB. Hence, let us assume that $MRM \approx MRB, MRV \approx MRW$. 

\begin{figure}[h]
\centering
\includegraphics[scale=0.25]{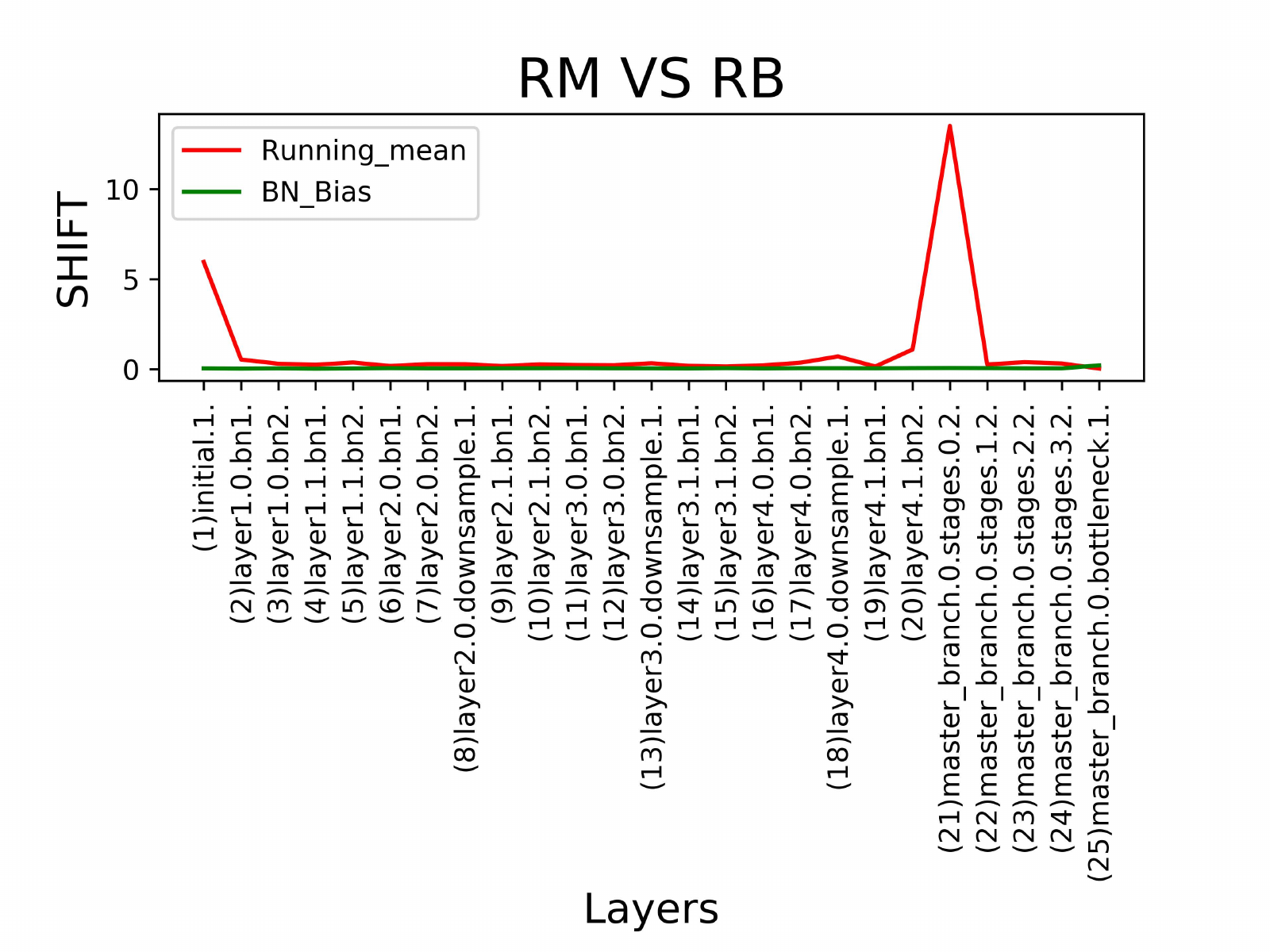}
\includegraphics[scale=0.25]{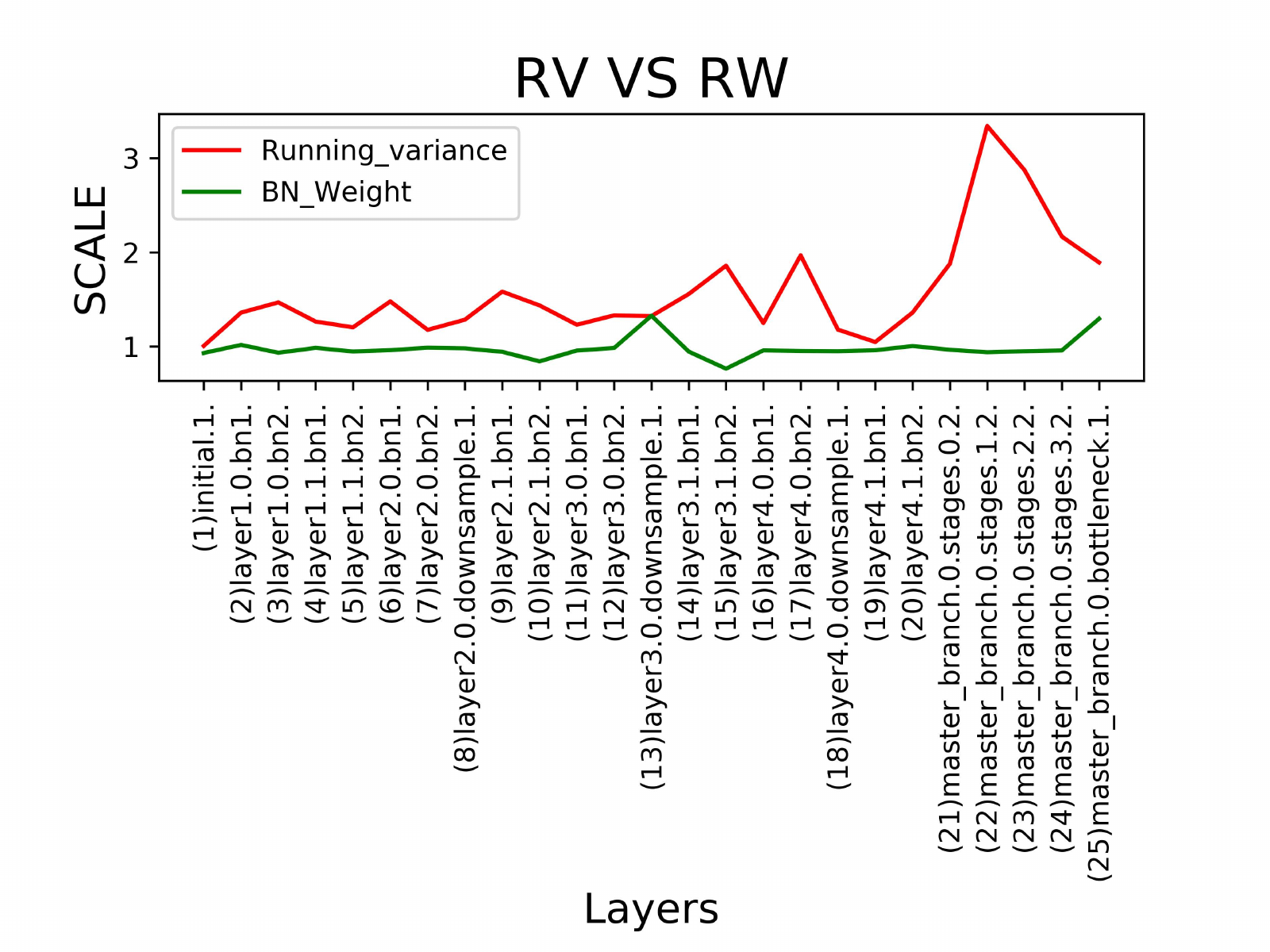}

\caption{PSPNet:Scale and shift in BN layers. Running Mean:MRM;BN Bias:MRB;Running variance:MRV;BN Weight:MRW. 
}
\label{pspnet-scale-shift}
\end{figure}

Plot the figure of $MRM, MRV, MRW, MRB$. Figure \ref{pspnet-scale-shift}.

The changes of MRM, MRV, MRW, and MRB basically conform to the above reasoning for Figure 7. In addition, UNet and PSPNet's RM are not reusable in the first layer.

\begin{figure}[h]
\centering
\includegraphics[scale=0.25]{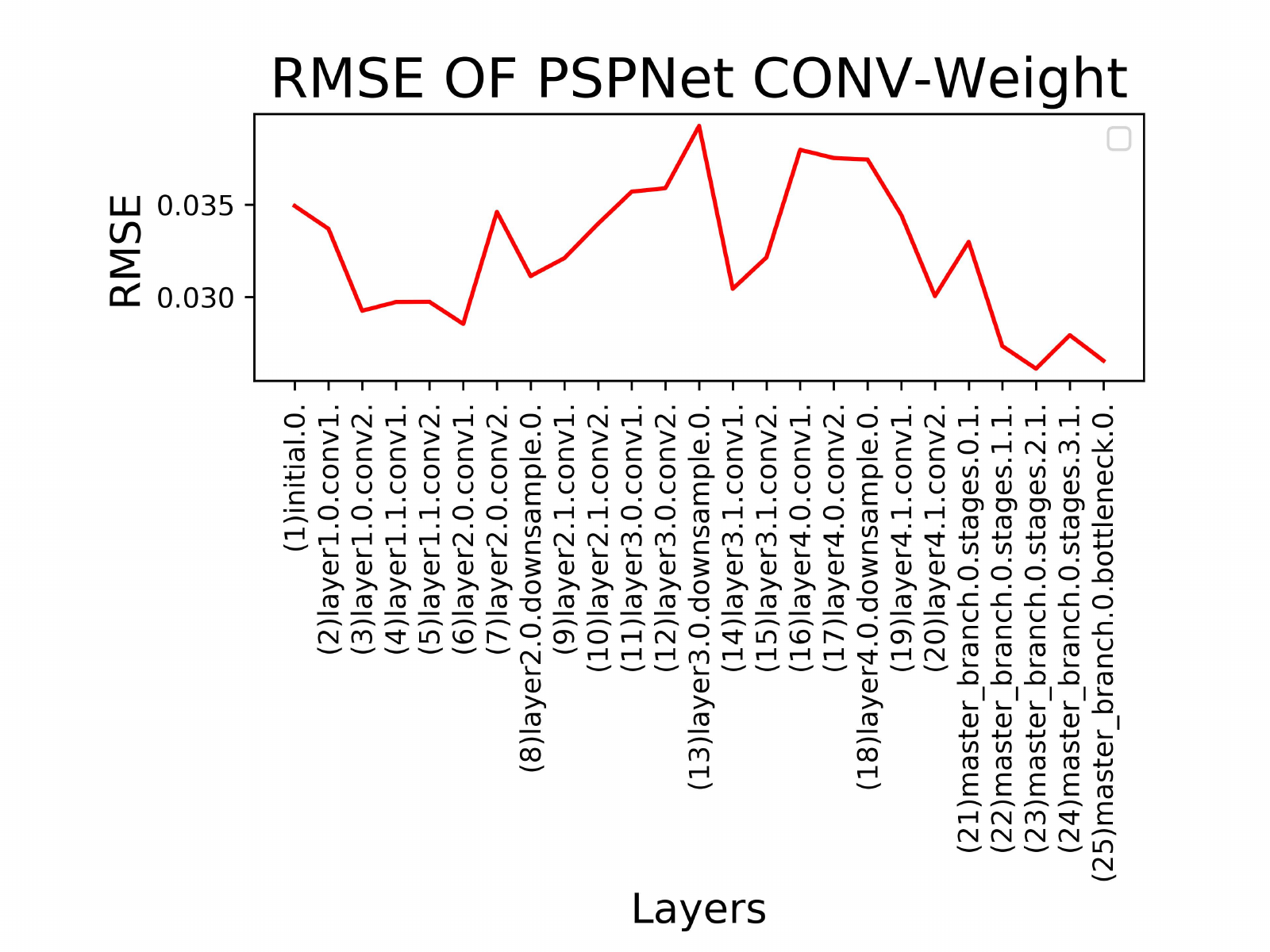}
\includegraphics[scale=0.25]{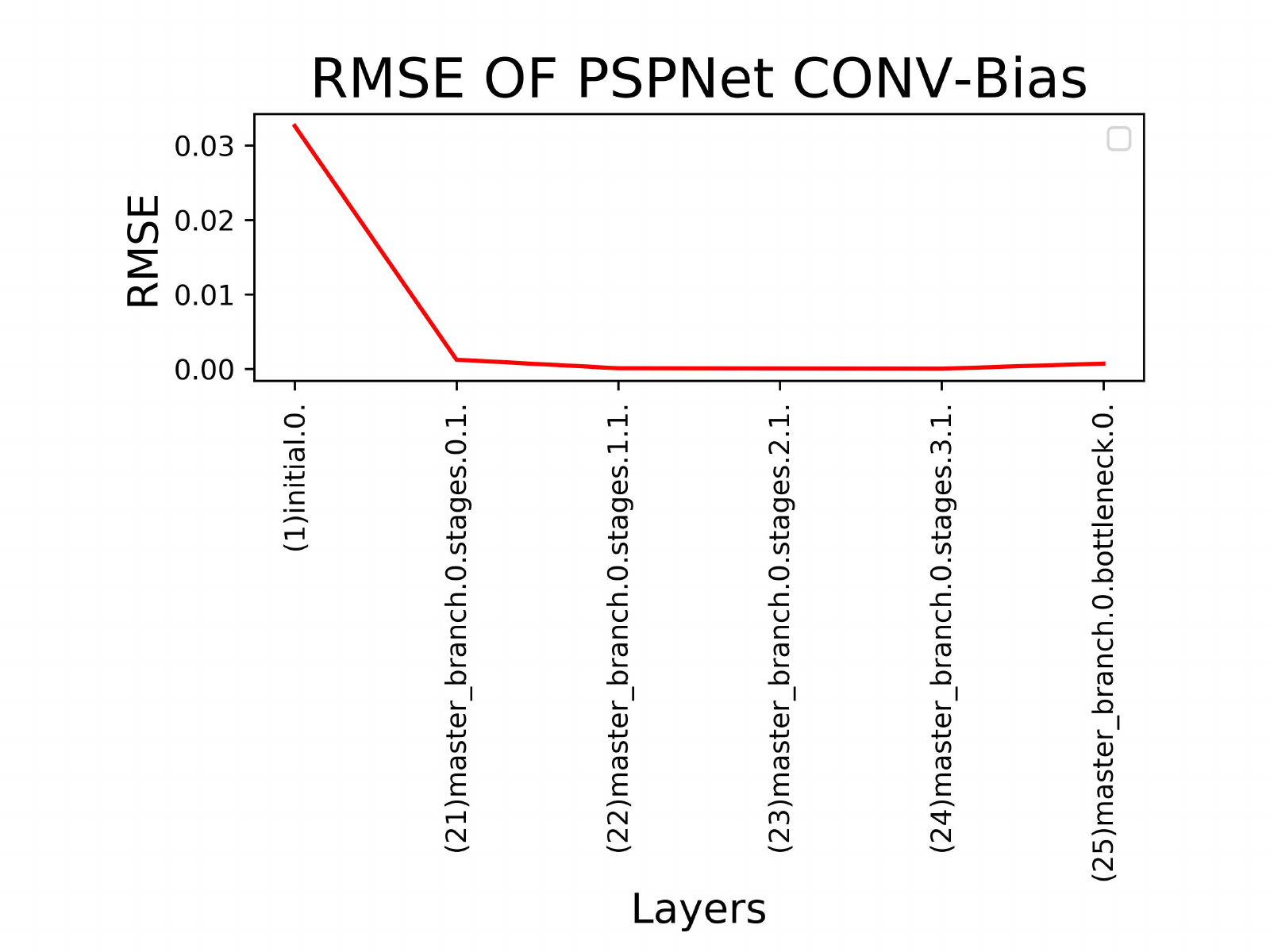}

\caption{PSPNet:The RMSE of W, B, which are the convolutional layers’ parameters, corresponding
to the image segmentation model and image auto-encoding model. 
}
\label{pspnet-RMSE}
\end{figure}

{\bfseries RMSE of convolution layer:} Figure \ref{pspnet-RMSE}

The RMSE of PSPNet's W and UNet's W are in the same order of magnitude, however PSPNet convolution kernel reuse indicates regular changes in Figure 8. Hence, B is reusable.

\subsubsection{1.3 Comparative summary between UNet and PSPNet}
According to the comparison between UNet and PSPNet parameter reuse results, it is found that the influence of parameter reuse on UNet is greater than that of PSPNet. In the BN layer, RM and RV have a greater influence on the result than RW and RB. The convolution kernels reuse results have huge differences, despite the RMSE of UNet being in the same order of magnitude as PSPNet. The result conforms to the reasoning given in Analysis of BN layer and convolution layer. 

\section{The RMSE of different tasks and different datasets in convolution layers}

Concerning network parameter reuse, the intuitive requirement is that the parameters in the network are generic. This paper maintains that it is also related to the difference of parameters. In order to explore the differences between network parameters of different models, this section calculates the RMSE between the segmentation and auto-encoder corresponding layer parameters of T1, RVSC, and ACDC datasets. 

\begin{figure}[h]
\centering

\includegraphics[scale=0.2]{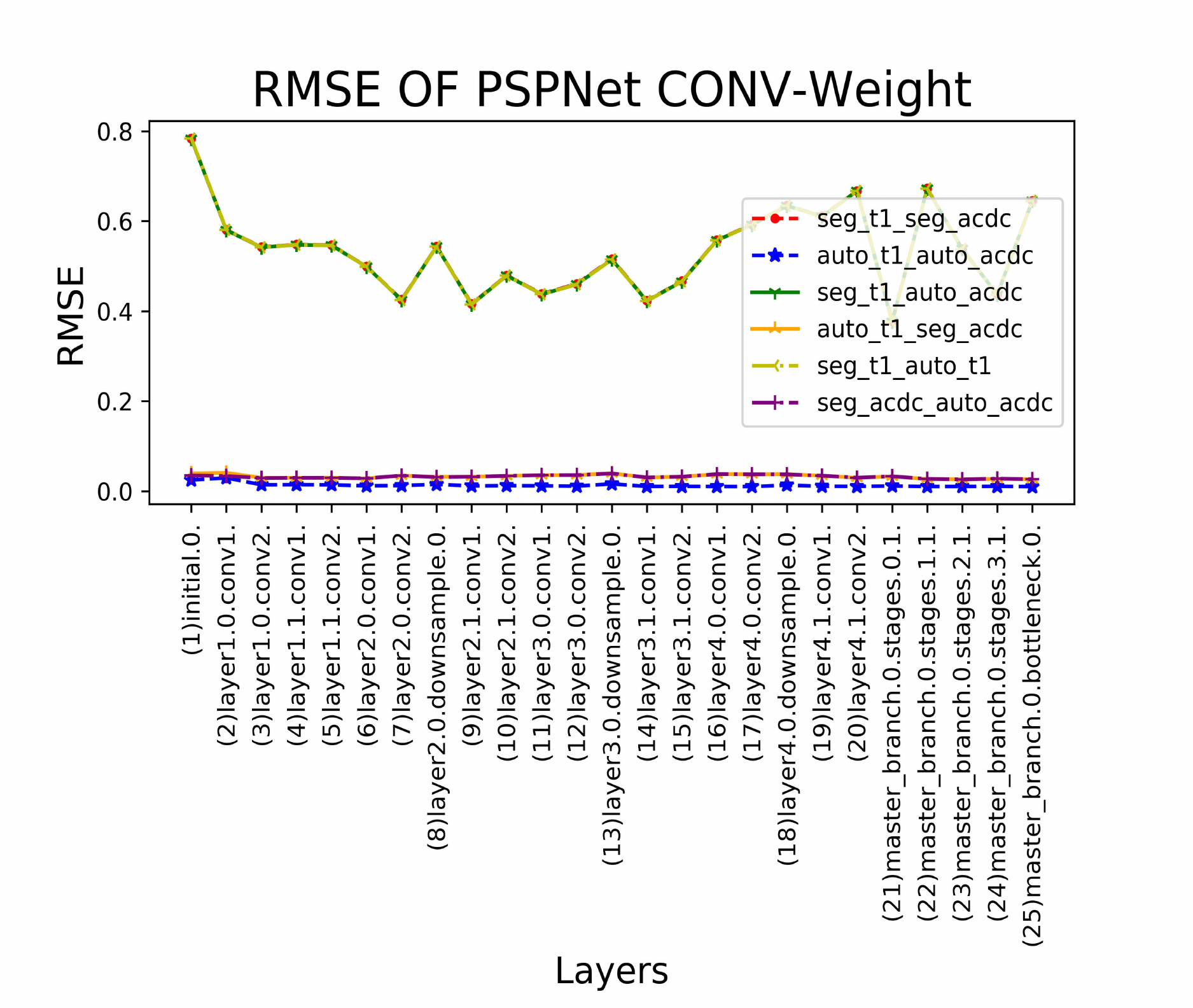}
\includegraphics[scale=0.2]{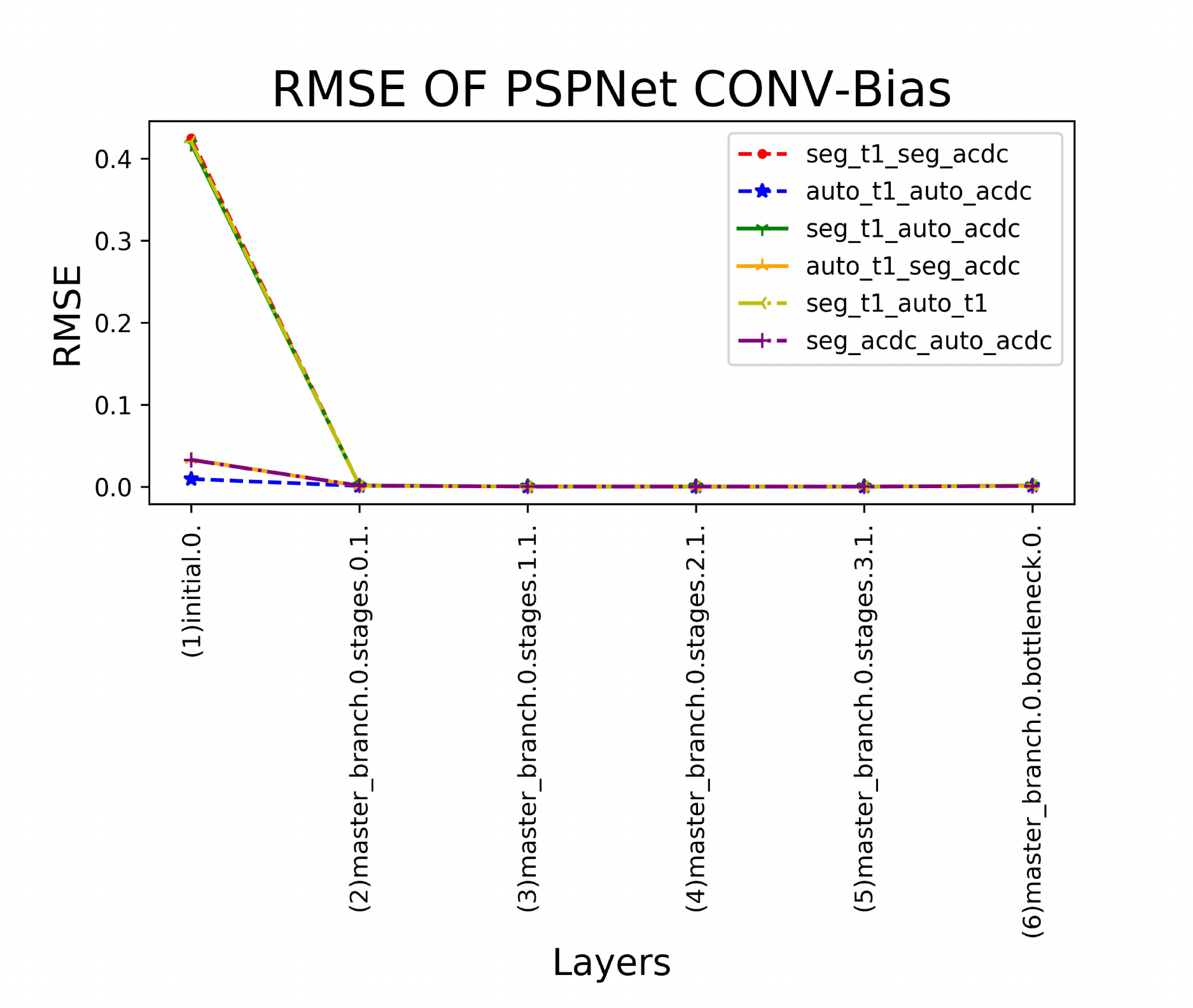}

\caption{Different models of PSPNet (RESnet-18) network RMSE. RMSE is calculated by two sets of models, seg\_T1, auto\_T1, seg\_ACDC and auto\_ACDC in the annotation, respectively denote the auto-encoder and segmentation model of T1 dataset and the auto-encoder and segmentation model of ACDC dataset. 
}

\label{pspnet-RMSE_T1_ACDC}
\end{figure} 
\begin{figure}[h]
\centering

\includegraphics[scale=0.2]{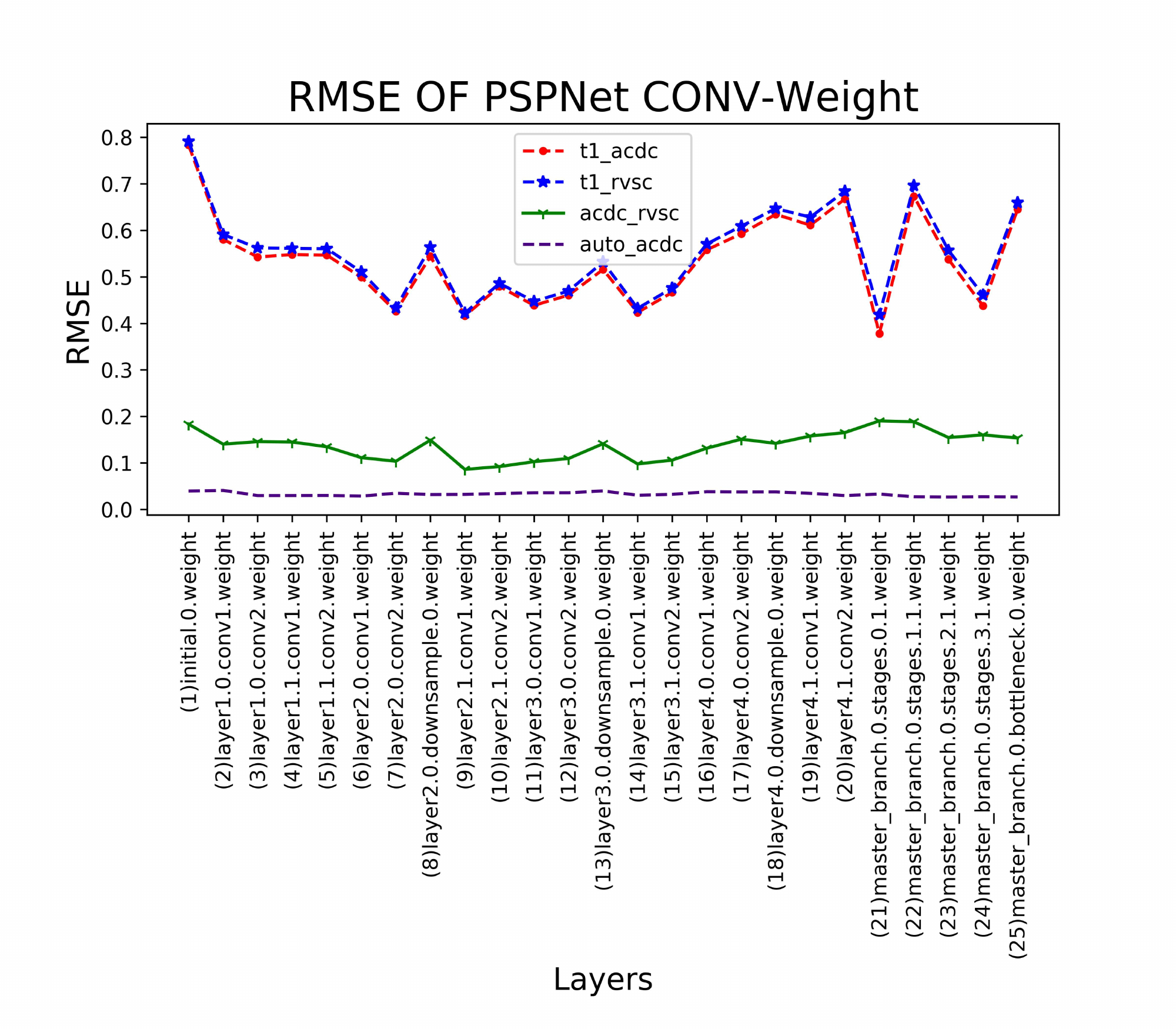}
\includegraphics[scale=0.2]{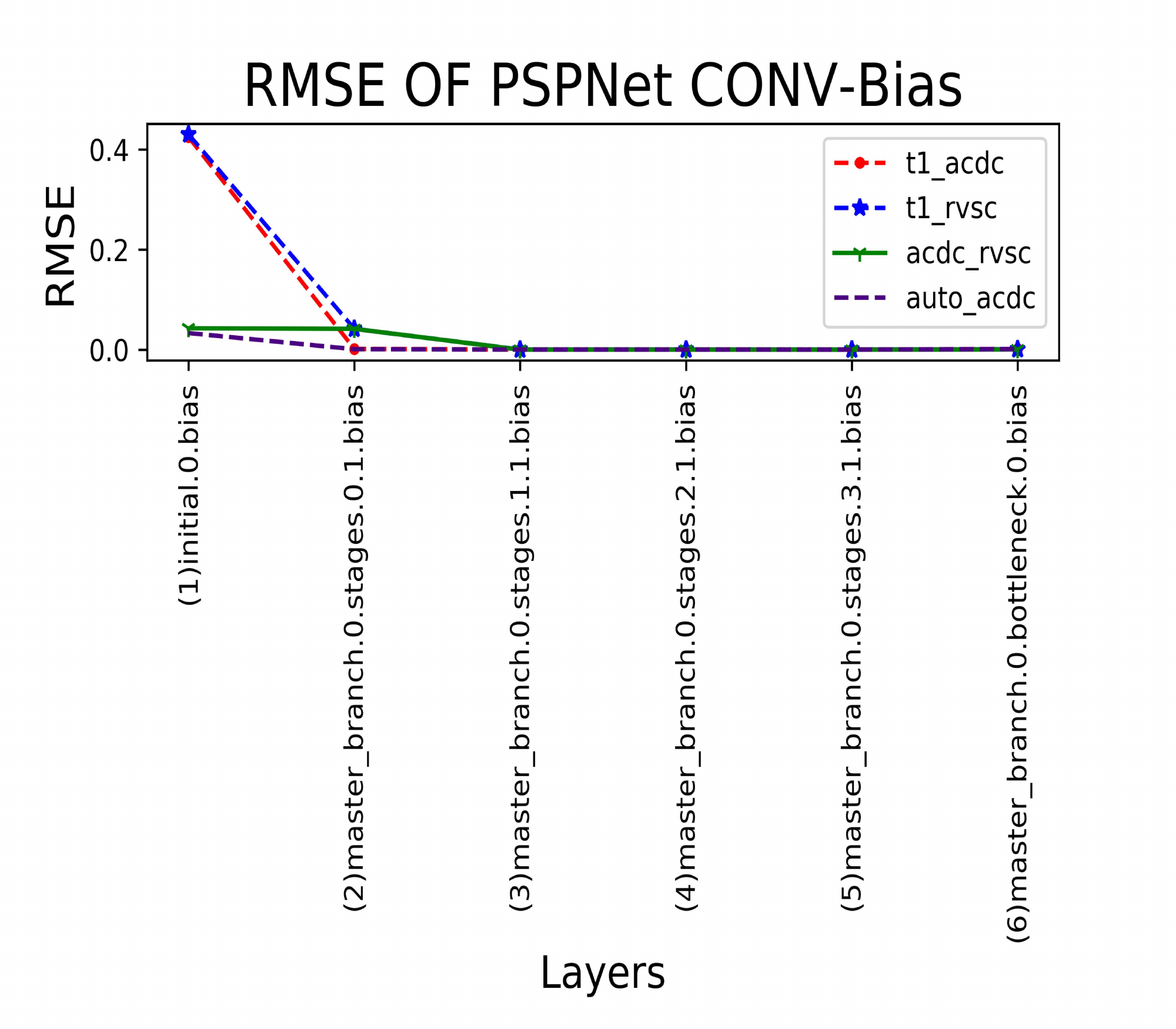}

\caption{RMSE is calculated by four models of PSPNet (RESnet-18) network. T1, RVSC, and ACDC in the annotation, respectively, denote the image segmentation model of T1, RVSC, and ACDC datasets. The image auto-encoder model of T1 dataset is denoted by auto. 
}

\label{pspnet-RMSE_T1_ACDC_RVSC}
\end{figure} 
RMSE images in Figure 11 were roughly divided into two groups, all the RMSE values with large differences were related to the T1 dataset segmentation model, while the others have small differences. Figure 12 images are roughly divided into three groups. RMSE values of corresponding layer parameters of the T1 image auto-encoder model and the ACDC image segmentation model exhibit minimal differences. 

The above phenomena can be summarized as follows:
A. Same dataset, different tasks (segmentation and image auto-encoder), and the RMSE magnitude is uncertain. 
B. Different datasets, different tasks, and the RMSE magnitude is uncertain. 
C. Different datasets, same task, and the RMSE magnitude is uncertain. 
According to the above results, there is a slight difference between the T1 dataset auto-encoder parameters and the ACDC dataset image segmentation parameters, hence it can be inferred that the T1 dataset image auto-encoder parameters can be reused in an ACDC segmentation task. 

\section{Parameter reuse experiment}~{}   

In order to verify the above conclusion about network parameter reuse, this section reuses the parameters learned from the T1 image auto-encoder, image segmentation, and RVSC image segmentation into an ACDC image segmentation. The experiment is divided into three parts. The first part was randomly initialized without loading any parameters. The second part loaded the T1 auto-encoder, image segmentation, and RVSC image segmentation parameters and then fixed the inferred reusable parameters to train. In the third part, the T1 auto-encoder, image segmentation, and RVSC image segmentation parameters were loaded and then used to fine-tune the inferred reusable parameters. 

\paragraph{3.1 Result of PSPNet}~{}  

According to the RMSE of the T1 auto-encoder and ACDC segmentation, it is speculated that the parameters which are non-reusable and grouped by layers of PSPNet (ResNet18) are RM: 1, 17; W: 1, 6, 11, 12, 16, 17, 19, 20; B: 1. All other parameters were reused. If each variable and each layer is treated independently , a total of 92\% of the parameters were reused in this experiment. 

\begin{table}[h]
    \centering
    \caption{Segmentation results of PSPNet (RESnet-18) network reuse T1 dataset auto-encoder and segmentation parameters on the ACDC dataset. }
    \scalebox{0.5}{
    \begin{tabular}{|l|l|l|l|l|l|}
        models & class-0 & class-1 & class-2 & class-3 & mean \\ 
        \multicolumn{6}{c}{10 samples}\\
        1. Don't load pretrained parameters & 0.99 & 0.62 & 0.60 & 0.70 & 0.73 \\ 
        2. T1\_seg$\rightarrow$ACDC\_seg freeze & 0.99 & 0.47 & 0.56 & 0.67 & 0.67 \\ 
        3. T1\_seg$\rightarrow$ACDC\_seg Do not freeze & 1.00 & 0.48 & 0.55 & 0.71 & 0.69 \\ 
        4. T1\_auto$\rightarrow$ACDC\_seg freeze & 0.99 & 0.70 & 0.70 & 0.84 & 0.81 \\ 
        5. T1\_auto$\rightarrow$ACDC\_seg Do not freeze & 1.00 & 0.75 & 0.75 & 0.88 & 0.85 \\ 
        6. RVSC\_seg$\rightarrow$ACDC\_seg freeze & 0.99 & 0.57 & 0.53 & 0.72 & 0.70 \\ 
        7. RVSC\_seg$\rightarrow$ACDC\_seg Do not freeze & 0.99 & 0.54 & 0.55 & 0.74 & 0.70 \\
        \multicolumn{6}{c}{50 samples}\\
        1. Don't load pretrained parameters & 1.00 & 0.81 & 0.82 & 0.90 & 0.88 \\ 
        2. T1\_seg$\rightarrow$ACDC\_seg freeze & 0.99 & 0.69 & 0.75 & 0.84 & 0.82 \\ 
        3. T1\_seg$\rightarrow$ACDC\_seg Do not freeze & 0.99 & 0.67 & 0.75 & 0.84 & 0.81 \\ 
        4. T1\_auto$\rightarrow$ACDC\_seg freeze & 1.00 & 0.84 & 0.83 & 0.90 & 0.89 \\ 
        5. T1\_auto$\rightarrow$ACDC\_seg Do not freeze & 1.00 & 0.86 & 0.86 & 0.93 & 0.91 \\ 
        6. RVSC\_seg$\rightarrow$ACDC\_seg freeze & 0.99 & 0.70 & 0.70 & 0. 86 & 0.72 \\ 
        7. RVSC\_seg$\rightarrow$ACDC\_seg Do not freeze & 0.99 & 0.67 & 0.74 & 0.82 & 0.81 \\ 
    \end{tabular}
    }
\end{table}

In the table, 10 and 50: The number of samples. In order to ensure the uniformity of the experiment, the validation set is the same people as the first part, and 10 people in the training set are randomly selected from 50 people in the original training set. Don not load pre-trained parameters: we use random initialization. Freeze: freeze the parameters that load the pre-trained parameters. Do not Freeze: the pre-trained parameters are trained with the network training. seg: segmentation parameters; auto: auto-coder parameters; A$\rightarrow$B: transfer A's parameters to B. The above results also explain why the results sometimes vary greatly when papers are reproduced. 

We can draw the conclusions from the above results, 1. Loading pre-training parameters may not necessarily improve the results. 10,50 sample: 1VS{2,3,6,7}. 2. According to the results presented in Figure 11, Figure 12, and 10, 50 sample: {4,5}VS{1,2,3,6,7} of this experiment, it can be concluded that a sufficiently small difference in convolutional kernels is beneficial to the results. 3. In practice, we can use a part of the target dataset to train a model for calculating RMSE with pre-trained convolution kernels for judging whether the given parameters are suitable for parameter reuse.  

According to Figures 11 and 12, we draw the conclusion that:  

a. Transfer learning results are not always satisfactory. It is required that the parameters of the original ideal model to be transferred should be similar enough to the parameters of the ideal model in this dataset. 

b. The essence of transfer learning lies in the similarity of transfer parameters to the ideal parameters, which is not derived from the similarity of dataset or task, and needs to be analyzed on a case-by-case basis. 

\section{Conclusion}

To summarize, this paper proposes an approach to explore the reusability of network parameters, which is verified theoretically and experimentally. Network parameters are reusable for two reasons:  

1. Networks have reusability. 

2. There is a slight difference between the pre-training parameters and the ideal network parameters. If the difference is large, it is not conducive to parameter reuse. 

According to the experiment in this study, the following conclusions can be drawn:  

1. PSPNet with short-connection structure (ResNet) is an RN, while UNet with long-Connection structure is a non-reusable network.
 
2. The order of magnitude of RMSE values of convolution kernels is different in different tasks and datasets, hence it needs to be analyzed concretely.

3. RM and RB change the data distribution by changing the data shift, thus affecting the network results. RV and RW affect the network results by changing data scaling and data distribution. RM and RV of the BN layer are more important to the final result than RW and RB, which are almost entirely reusable. 

%4. RM of BN layers have a great influence on the final result in the first layer. 

4. Transfer learning may not necessarily improve the results. The result of transfer learning depends on the reusability of the network and the difference of the convolution kernel. The same is true for semi-supervised learning.

\bibliography{refs}
\bibliographystyle{unsrtnat}

\end{document}